\documentclass[12pt]{article}

\usepackage[a4paper,text={16.8cm,22.4cm}]{geometry}
\usepackage{amsmath,amsfonts,slashed,amssymb,tikz,bm,psfrag,graphicx,color,dsfont}
\usepackage{multicol}
\usepackage{auto-pst-pdf}

\usetikzlibrary{decorations.pathmorphing}
\usetikzlibrary{decorations.markings}
\usetikzlibrary{arrows,shapes}

\RequirePackage[sort&compress,square,comma,numbers]{natbib}
\allowdisplaybreaks 
\newcommand{\ord}{{\cal O}}
\newcommand{\e}{\epsilon}
\newcommand{\tx}{\text}
\newcommand{\pa}{\partial}
\newcommand{\La}{\mathcal{L}}
\newcommand{\tr}{\tx{Tr}}
\newcommand{\TeV}{\tx{TeV}}
\newcommand{\SM}{\tx{SM}}
\newcommand{\Xb}{\boldsymbol{X}} 
\newcommand{\mkk}{M_{\text{KK}}} 
\newcommand{\al}{\alpha}
\newcommand{\be}{\beta}
\newcommand{\ga}{\gamma}
\newcommand{\pEh}{{\hat p_E}}

\tikzset{
vector/.style={decorate, decoration={snake,amplitude=2pt}, draw,segment length=6pt},
fermion/.style={draw=black, postaction={decorate}, decoration={markings,mark=at position .55 with {\arrow[draw=black]{>}}}},
fermionbar/.style={draw=black, postaction={decorate}, decoration={markings,mark=at position .55 with {\arrow[draw=black]{<}}}},
fermionnoarrow/.style={draw=black},
scalararrow/.style={dashed,draw=black, postaction={decorate}, decoration={markings,mark=at position .55 with {\arrow[draw=black]{>}}}},
scalar/.style={dashed}, 
gluon/.style={decorate, draw=black, decoration={coil,amplitude=4pt, mirror,segment length=7.85pt,aspect=0.8}},
vertex/.style={draw,circle,fill=black,inner sep=0pt,minimum size=1.2mm},  
comp/.style={line width=5pt, draw=blue!20},
elem/.style={line width=1pt, draw=black},
comp2/.style={line width=5pt, draw=blue!20},
elem2/.style={line width=1pt, draw=black},
ghost/.style={dotted,draw=black}
}

\begin{document}

\begin{titlepage}

\begin{flushright}
\normalsize
MITP/13-87\\
December 20, 2013 \\
arXiv:1312.5731
\end{flushright}

\vspace{0.1cm}
\begin{center}
\Large\bf
Higgs Decay into Two Photons in a Warped\\
Extra Dimension 
\end{center}

\vspace{0.5cm}
\begin{center}
Juliane Hahn$^a$, Clara H\"orner$^a$, Raoul Malm$^a$, Matthias Neubert$^{a,b}$,\\ 
Kristiane Novotny$^a$ and Christoph Schmell$^a$\\
\vspace{0.7cm} 
{\sl ${}^a$PRISMA Cluster of Excellence \& Mainz Institute for Theoretical Physics\\
Johannes Gutenberg University, 55099 Mainz, Germany\\[3mm]
${}^b$Department of Physics, LEPP, Cornell University, Ithaca, NY 14853, U.S.A.}
\end{center}

\vspace{0.8cm}
\begin{abstract}
A detailed five-dimensional calculation of the Higgs-boson decay into two photons is performed in both the minimal and the custodially protected Randall-Sundrum (RS) model, where the Standard Model (SM) fields propagate in the bulk and the scalar sector lives on or near the IR brane. It is explicitly shown that the $R_\xi$ gauge invariance of the sum of diagrams involving bosonic fields in the SM also applies to the case of these RS scenarios. An exact expression for the $h\to\ga\ga$ amplitude in terms of the five-dimensional (5D) gauge-boson and fermion propagators is presented, which includes the full dependence on the Higgs-boson mass. Closed expressions for the 5D $W$-boson propagators in the minimal and the custodial RS model are derived, which are valid to all orders in $v^2/\mkk^2$. In contrast to the fermion case, the result for the bosonic contributions to the $h\to\ga\ga$ amplitude is insensitive to the details of the localization of the Higgs profile on or near the IR brane. The various RS predictions for the rate of the $pp\to h\to\ga\ga$ process are compared with the latest LHC data, and exclusion regions for the RS model parameters are derived. 
\end{abstract}
\vfil

\end{titlepage}

\section{Introduction}

After the discovery of the Higgs boson in July 2012 \cite{ATLAS:2012gk,CMS:2012gu}, a solution to the hierarchy problem -- the question why the electroweak scale is so much lower than the Planck scale -- is more urgently needed than ever. Among the numerous possibilities to solve the hierarchy problem, the most popular approach is low-scale supersymmetry. A promising alternative is given by models with a warped extra dimension \cite{Randall:1999ee}, in which the Standard Model (SM) is embedded in a slice of anti-de Sitter space, while the Higgs field is localized on the ``infra-red (IR) brane'', one of the two four-dimensional hyper-manifolds that bound the extra dimension. These so-called Randall-Sundrum (RS) models can provide a natural explanation for this vast hierarchy, since the fundamental ultra-violet (UV) cutoff is the warped Planck scale, which near the IR brane takes values in the TeV range. Moreover, by allowing the fermion fields to propagate into the bulk, warped extra-dimension models can also provide an explanation for the hierarchies observed in the flavor sector \cite{Grossman:1999ra,Gherghetta:2000qt,Huber:2000ie} and the smallness of flavor-changing neutral currents \cite{Agashe:2004ay,Agashe:2004cp,Csaki:2008zd,Casagrande:2008hr,Blanke:2008zb,Blanke:2008yr,Bauer:2009cf}.

Unfortunately, none of the Kaluza-Klein (KK) particles predicted in extra-dimensional extensions of the SM have been observed yet, and electroweak precision measurements indicate that these particles are probably too massive for a direct detection at the LHC. Thus, indirect searches like the precision measurements of the Higgs-boson couplings to SM particles, which are accessible via studies of both the Higgs production cross section and its decay rates, become more and more attractive. Concerning warped extra dimensions, especially the loop-mediated Higgs couplings to gluons and photons could give hints about the existence of additional KK particles. While the gluon-fusion process has been discussed extensively in several works \cite{Djouadi:2007fm,Falkowski:2007hz,Cacciapaglia:2009ky,Bhattacharyya:2009nb,Bouchart:2009vq,Casagrande:2010si,Azatov:2010pf,Azatov:2011qy,Goertz:2011hj,Carena:2012fk,Malm:2013jia}, the present paper focuses on the Higgs decay into two photons, which was investigated in \cite{Cacciapaglia:2009ky,Bhattacharyya:2009nb,Bouchart:2009vq,Casagrande:2010si,Azatov:2010pf}. The first analysis of the effects of the KK tower of the $W$ boson on the $h\to\ga\ga$ amplitude was performed in \cite{Bouchart:2009vq}. The first complete calculation of the $h\to\gamma\gamma$ decay rate, in which both the Yukawa couplings to the $Z_2$-even and $Z_2$-odd fermions were included, was performed in \cite{Casagrande:2010si}. It was found in this paper that the Higgs decay rate into two photons is enhanced relative to the SM due to the effect of the KK fermions, which turned out to give the dominating correction. At about the same time, an independent analysis of the $h\to\gamma\gamma$ decay rate came to the opposite conclusion \cite{Azatov:2010pf}. An explanation for these deviating results was first given in \cite{Carena:2012fk}, followed by a five-dimensional (5D) analysis in \cite{Malm:2013jia}, which found that the two results belong to two different scenarios of the RS model.\footnote{These papers analyzed the quark KK-tower contribution to the Higgs production process $gg\to h$. Up to different factors for the color multiplicity and electric charges, an analogous discussion holds for the quark and lepton KK-tower contributions to $h\to\gamma\gamma$.}  
While these papers focused on the contribution stemming from the fermionic KK states, in the present work we provide a detailed analysis of the bosonic loop contributions to the $h\to\gamma\gamma$ amplitude, which in unitary gauge stem from the $W$ bosons and their KK excitations. The advantages of this approach are that we are able to derive an exact result, which includes the full dependence on the Higgs-boson mass and holds to all orders in $v^2/\mkk^2$, where $v$ is the Higgs vacuum expectation value (vev), and $M_{\rm KK}$ sets the mass scale for the low-lying KK excitations of the SM particles. It is also straightforward to extend our formulas to the case where the Higgs boson lives in the bulk of the extra dimension. Our approach also allows us to carefully study the effects of the fifth components of the gauge fields, whose profiles are discontinuous on the IR brane, similar to the $Z_2$-odd fermion profiles which indeed require a careful treatment.

Our paper is structured as follows: In Section~\ref{sec:pre} we define the set-up and comment on the necessity to distinguish between the so-called brane-Higgs and narrow bulk-Higgs scenarios when calculating the fermionic loop contributions to the $gg\to h$ and $h\to\ga\ga$ amplitudes. In Section~\ref{sec:analysis} we discuss the general structure of the $h\to\gamma\gamma$ amplitude and summarize the results for the fermionic contributions from charged quarks and leptons propagating in the loop. We then focus on the bosonic loop contributions, calculate them in the KK-decomposed theory and show that the result for the contributions of each individual KK mode is gauge invariant. In the next step we resum the KK towers and derive an exact formula for the $h\to\gamma\gamma$ amplitude in terms of an overlap integral over the Higgs-boson profile and the transverse part of the 5D gauge-boson propagator, including the exact dependence on the Higgs-boson mass. To the best of our knowledge, such a formula has not been obtained before. We derive an explicit, closed expression for the 5D propagator and show that the overlap integral is insensitive to the precise details of shape of the Higgs-boson profile, once this profile is localized very close to the IR brane. By expanding our results in powers of $v^2/\mkk^2$, we can identify the contributions from the $W$ bosons (with modified couplings to the Higgs boson) and their KK towers, confirming the results of \cite{Bouchart:2009vq}. In Section~\ref{sec:custodial} we generalize our findings to an extended version of the RS model with a custodial symmetry protecting electroweak precision observables \cite{Agashe:2003zs,Csaki:2003zu,Agashe:2006at}. Again, we obtain an exact formula for the $h\to\gamma\gamma$ amplitude and, for the first time, for the 5D gauge-boson propagator in the custodial RS model. When expanded to order $v^2/\mkk^2$, our findings for the contributions of the $W$ boson and its KK excitations are consistent with the findings of \cite{Casagrande:2010si}. Phenomenological implications of our results in the context of the latest LHC data are discussed in Section \ref{sec:pheno}, where we study the Higgs decay into two photons in two different versions of the minimal and the custodially protected RS model. We illustrate the magnitude of the effects as a function of the mass of the lightest KK gluon state and the scale of the 5D Yukawa couplings, and derive the regions in parameter space that are already excluded by recent LHC measurements. Our main results are summarized in the conclusions.

\section{Preliminaries}
\label{sec:pre}

We focus on RS models where the electroweak symmetry-breaking sector is localized on or near the IR brane. The extra dimension is chosen to be an $S^1/Z_2$ orbifold parametrized by a coordinate $\phi \in [-\pi, \pi]$, with two branes localized on the orbifold fixed-points at $\phi=0$ (UV brane) and $|\phi|=\pi$ (IR brane). The RS metric reads \cite{Randall:1999ee}
\begin{equation}
\label{eqn:RSmetric}
   ds^2 = e^{-2\sigma(\phi)}\,\eta_{\mu\nu}\,dx^\mu dx^\nu - r^2 d\phi^2
   = \frac{\epsilon^2}{t^2} \left( \eta_{\mu\nu}\,dx^\mu dx^\nu
    - \frac{1}{M_{\rm KK}^2}\,dt^2 \right) ,
\end{equation}
where $e^{-\sigma(\phi)}$ with $\sigma(\phi)=kr|\phi|$ is referred to as the warp factor, and the size $r$ and curvature $k$ of the extra dimension are assumed to be of Planck size, $k\sim 1/r\sim M_{\rm Pl}$. The quantity $L=\sigma(\pi)=kr\pi$ measures the size of the extra dimension and is chosen to be $L\approx 33-34$ in order to explain the hierarchy between the Planck and the TeV scales. With the help of the curvature $k$ and the warp factor evaluated at the IR brane, $\epsilon=e^{-\sigma(\pi)}$, one defines the KK mass scale as $M_{\rm KK}=k\epsilon$. On the right-hand side of \eqref{eqn:RSmetric} we have introduced the dimensionless coordinate $t$ defined by $t=\epsilon\,e^{\sigma(\phi)}\in[\epsilon,1]$,\footnote{This variable is related to the frequently used conformal coordinate $z$ by the rescaling $z=t/M_{\rm KK}$.}  
which will be used throughout this work.

Our paper deals with the minimal RS model based on the SM gauge group, as well as with an extended RS model with a larger bulk gauge group, which after electroweak symmetry breaking includes an $SU(2)$ custodial symmetry protecting the $T$ parameter and suppressing larger corrections to the $Z\bar b b$ couplings \cite{Agashe:2003zs,Csaki:2003zu,Agashe:2006at}. In both versions of the RS model, all fermions and gauge bosons are allowed to propagate into the bulk, resulting in infinite towers of heavy KK copies of the SM particles. In contrast, the scalar sector is assumed to reside on or near the IR brane so as to provide a solution to the hierarchy problem. We recall that RS models are effective field theories valid up to a position-dependent UV cutoff \cite{Randall:2001gb,Pomarol:2000hp,Choi:2002wx,Goldberger:2002cz} 
\begin{equation}
   \Lambda_{\rm UV}(t)\approx M_{\rm Pl}\,e^{-\sigma(\phi)} 
   = M_{\rm Pl}\,\frac{\e}{t}
   \equiv \frac{\Lambda_{\rm TeV}}{t} \,.
\end{equation}
At this scale, gravity becomes strong and the model needs to be UV-completed into a theory of quantum gravity. The fact that RS models are defined with an inherent UV cutoff can be used to distinguish between different scenarios for the localization of the Higgs sector. We shall consider a scalar field $\Phi$ localized on or very near the IR brane. Its profile along the extra dimension is described by a normalized distribution $\delta^\eta(t-1)$ with width $\eta\ll 1$. For $\eta\to 0$ this profile becomes a $\delta$-function, corresponding to a strictly brane-localized scalar field. Following \cite{Malm:2013jia}, we introduce the characteristic width of the Higgs field along the extra dimension as $\Delta_h^{-1}\equiv\eta/v$. If the inverse width is larger than the inherent UV cutoff near the IR brane, $\Delta_h\gg\Lambda_{\rm TeV}$, then the Higgs profile cannot be resolved by the high-momentum modes contained in the effective Lagrangian of the RS model, and hence for all practical purposes such a scalar field can be regarded as a {\em brane-localized Higgs field}. On the other hand, if the inverse Higgs width is smaller than the cutoff scale, then the modes of the effective theory can resolve its profile, and we speak of a {\em bulk Higgs field}. While calculations in generic bulk-Higgs models are rather complicated, it has been shown in \cite{Malm:2013jia,Delaunay:2012cz} that there is the possibility of obtaining analytical results for the special case of a {\em narrow bulk-Higgs field}, whose inverse width is such that $M_{\rm KK}\ll\Delta_h\ll\Lambda_{\rm TeV}$. As has been explained in \cite{Carena:2012fk,Malm:2013jia}, the fermionic loop contributions to the $gg\to h$ and $h\to\ga\ga$ amplitudes are sensitive to the details of the localization mechanism, and indeed the results obtained in a brane-localized Higgs scenario \cite{Casagrande:2010si,Goertz:2011hj} differ significantly from those derived for a narrow bulk-Higgs field \cite{Azatov:2010pf,Frank:2013un}. We thus need to distinguish between the two types of localization mechanisms in our phenomenological analysis. As we will show, however, the bosonic contributions to the $h\to\ga\ga$ amplitude are insensitive to the precise localization of the scalar sector and approach an unambiguous result in the limit where $\eta\ll 1$.

\section{\boldmath 5D analysis of the $h\to\ga\ga$ amplitude}
\label{sec:analysis}

Our goal is to calculate the $h\to\gamma\gamma$ decay amplitude entirely in terms of the 5D propagators for both gauge bosons and fermions. While the contributions from quarks and charged leptons can be easily deduced from the corresponding results for the $gg\to h$ amplitude, a detailed consideration of the gauge-boson contribution has not yet been performed in 5D language. Our approach in the present work is the following: In a first step, we calculate the bosonic contributions to the $h\to\gamma\gamma$ amplitude in the KK-decomposed, 4D effective theory and show that at each KK level the sum of all diagrams is gauge-invariant. The only contributing diagrams in unitary gauge are those with vector bosons propagating in the loop. We can then rewrite the amplitude, summed over KK states, as an expression involving the 5D gauge-boson propagator in the mixed momentum-position representation \cite{Randall:2001gb,Puchwein:2003jq,Contino:2004vy,Carena:2004zn,Csaki:2010aj}. We show that in the limit of a very narrow Higgs profile the amplitude approaches an unambiguous value, which is insensitive to the details of the Higgs localization mechanism. At the end of this section, we employ our exact results to derive expressions for the contributions of the zero modes (the standard $W$ bosons) and their infinite towers of KK excitations to the $h\to\ga\ga$ amplitude. 

We begin with the calculation in the minimal RS model with the SM gauge group in the bulk, broken to $U(1)_{\rm EM}$ on the IR brane, where the Higgs field develops a vev. Details for the implementation of the Higgs, gauge-boson, and gauge-fixing sectors in the context of this model and using our notations have been given in \cite{Casagrande:2008hr}, while Appendix~\ref{app:FR} includes a summary of the relevant Feynman rules needed for our analysis. Here, it suffices to mention that we decompose the 5D $W$-boson field into 4D mass eigenstates
\begin{equation}
\begin{aligned}\label{eqn:KKdecomM}
   W_\mu^\pm(x,t) &= \frac{1}{\sqrt{r}} \sum_{n=0}^\infty {\chi}_n^{\,W}(t)\,
    W_\mu^{\pm(n)}(x) \,, \\
   W_\phi^\pm(x,t) &= - \frac{1}{\sqrt{r}}\,\frac{L}{\pi} \sum_{n=0}^\infty
    \frac{1}{m_n^W}\,t\,\partial_t\chi_n^W(t)\,\varphi_W^{\pm(n)}(x) \,,
\end{aligned}
\end{equation}
where $W_\mu^{\pm(n)}$ are the KK modes of the $W$ bosons with masses $m_n^W$. The scalar particles $\varphi_W^{\pm(n)}$ are ``unphysical'' in the sense that they provide the longitudinal degrees of freedom of the $W$ bosons ($n=0$) and the massive $W$-boson KK modes ($n\ge 1$), and thus they can be gauged away. Indeed, the scalar fields $W_\phi^\pm$ mix with the charged Goldstone bosons arising from the Higgs sector. Assuming for the time being that the scalar sector is localized on the IR brane, we parameterize the Higgs doublet after electroweak symmetry breaking in the usual form
\begin{equation}
   \Phi(x) = \frac{1}{\sqrt{2}} 
   \begin{pmatrix}
    -i\sqrt{2} \varphi^+(x) \\
    v + h(x) + i\varphi^3(x) &	
   \end{pmatrix} .
\end{equation}
Throughout this paper, $v$ denotes the Higgs vev in the RS model, which differs from the SM value $v_\SM\approx 246\,\tx{GeV}$ by a small amount \cite{Bouchart:2009vq}. We determine $v$ from the shift to the Fermi constant derived in the RS model by considering (at tree level) the effect of the exchange of the infinite tower of KK gauge bosons on the rate for muon decay, using the definition $v_{\rm SM}=(\sqrt2\,G_F)^{-1/2}$. This yields \cite{Malm:2013jia}
\begin{equation}\label{kappav}
   \kappa_v = \frac{v}{v_{\rm SM}} = 1 + \frac{L m_W^2}{4c_\vartheta^2\mkk^2}
    + \ord\bigg( \frac{v^4}{\mkk^4} \bigg) \,,
\end{equation}
where $c_\vartheta=1$ in the minimal RS model. When we generalize our analysis to the case of an extended RS model with a custodial symmetry in Section~\ref{sec:custodial}, this relation will still hold, but the parameter $c_\vartheta$ will then take on a non-trivial value. The decomposition of the scalar fields $\varphi^\pm$ into the mass eigenstates $\varphi_W^{\pm(n)}$ reads~\cite{Casagrande:2008hr}
\begin{equation}\label{varphidecomp}
   \varphi^\pm(x) = \sum_{n=0}^\infty \frac{\tilde m_W}{m_n^W}\,
    \sqrt{2\pi}\,\chi_n^W(1)\,\varphi_W^{\pm(n)}(x) \,, \qquad
   \tilde m_W = \frac{g_5}{\sqrt{2\pi r}}\,\frac{v}{2} \,,
\end{equation}
where $\tilde m_W$ is the leading contribution to the $W$-boson mass in an expansion in powers of $v^2/\mkk^2$. The relation between the two parameters can be written as \cite{Casagrande:2008hr}
\begin{equation}\label{Wmassrela}
   \tilde m_W^2 = m_W^2 \left[ 1 + \frac{m_W^2}{2\mkk^2}
    \left( \frac{L}{c_\vartheta^2} - 1 + \frac{1}{2L} \right) 
    + \ord\bigg( \frac{v^4}{\mkk^4} \bigg) \right] ,
\end{equation}
where once again $c_\vartheta=1$ in the minimal RS model. Since the profile of the zero mode is flat up to corrections of order $v^2/\mkk^2$, it follows that 
\begin{equation}
   \sqrt{2\pi}\,\chi_0^W(1) = 1 - \frac{m_W^2}{2\mkk^2}
    \left( L - 1 + \frac{1}{2L} \right) + \ord\bigg( \frac{v^4}{\mkk^4} \bigg)
\end{equation}
is close to~1, and hence the fields $\varphi^\pm$ coincide with $\varphi_W^{\pm(0)}$ to leading order. Mixing effects arise at order $v^2/\mkk^2$ and higher. Note also that one can adjust the gauge-fixing Lagrangian so as to cancel any mixings between $W_\mu^\pm$ and the scalar fields $W_\phi^\pm$ and $\varphi^\pm$ \cite{Casagrande:2008hr}.

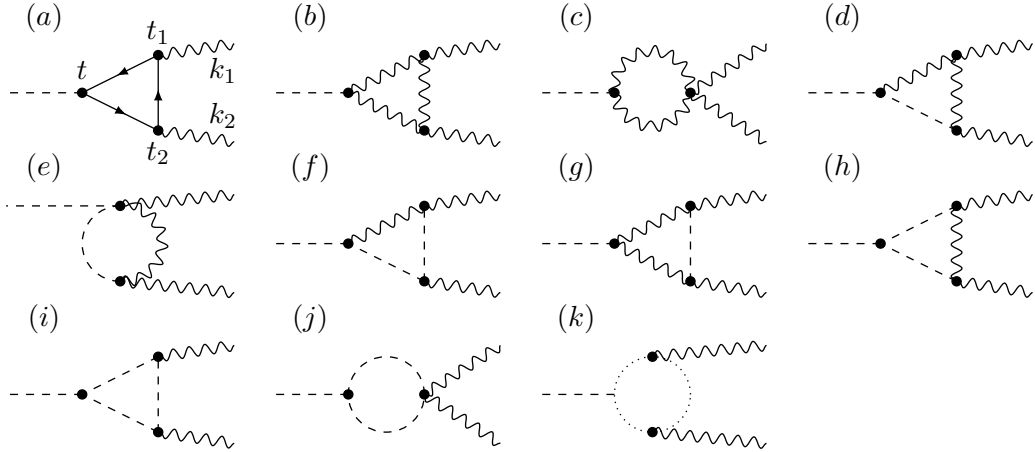
\begin{figure}[t!]
\begin{center}
\begin{tikzpicture}[line width=0.5pt,>=latex,scale=0.5]
\node (origin) at (0,0) {};
\begin{scope}[xshift=-7cm, yshift=0]
\draw[fermion] (1,1) -- (-1,0);
\draw[fermionbar] (1,-1) -- (-1,0);
\draw[fermion] (1,-1) -- (1,1);
\draw[scalar] (-1,0) -- (-3,0);
\draw[vector] (3,1.3) -- (1,1);
\draw[vector] (3,-1.3) -- (1,-1);
\draw (-1,0) node [vertex] {};
\draw (1,1) node [vertex] {};
\draw (1,-1) node [vertex] {};
\draw (-2,2) node {\small $(a)$};
\draw (-1,0.6) node {\small $t$};
\draw (1,1.6) node {\small $t_1$};
\draw (1,-1.6) node {\small $t_2$};
\draw (2.7,0.6) node {\small $k_1$};
\draw (2.7,-0.6) node {\small $k_2$};
\end{scope}
\begin{scope}[xshift=0cm, yshift=0]
\draw[vector] (1,1) -- (-1,0);
\draw[vector] (1,-1) -- (-1,0);
\draw[vector] (1,-1) -- (1,1);
\draw[scalar] (-1,0) -- (-3,0);
\draw[vector] (3,1.3) -- (1,1);
\draw[vector] (3,-1.3) -- (1,-1);
\draw (-1,0) node [vertex] {};
\draw (1,1) node [vertex] {};
\draw (1,-1) node [vertex] {};
\draw (-2,2) node {\small $(b)$};
\end{scope}
\begin{scope}[xshift=7cm, yshift=0]
\draw[scalar] (-1,0) -- (-3,0);
\draw[vector] (3,1.3) -- (1,0);
\draw[vector] (3,-1.3) -- (1,0);
\draw (-1,0) node  [vertex] {};
\draw (1,0) node  [vertex] {};
\draw (-2,2) node {\small $(c)$};
\draw[vector] (1,0) arc (0:360:1);
\end{scope}
\begin{scope}[xshift=14cm, yshift=0]
\draw[vector] (1,1) -- (-1,0);
\draw[scalar] (1,-1) -- (-1,0);
\draw[vector] (1,-1) -- (1,1);
\draw[scalar] (-1,0) -- (-3,0);
\draw[vector] (3,1.3) -- (1,1);
\draw[vector] (3,-1.3) -- (1,-1);
\draw (-1,0) node [vertex] {};
\draw (1,1) node  [vertex] {};
\draw (1,-1) node [vertex] {};
\draw (-2,2) node {\small $(d)$};
\end{scope}
\begin{scope}[xshift=-7cm, yshift=-4cm]
\draw[scalar] (0,1) -- (-3,1);
\draw[vector] (3,1.3) -- (0,1);
\draw[vector] (3,-1.3) -- (0,-1);
\draw (0,1) node  [vertex] {};
\draw (0,-1) node  [vertex] {};
\draw[vector] (0,-1) arc (-90:90:1);
\draw[scalar] (0,-1) arc (270:90:1);
\draw (-2,2) node {\small $(e)$};
\end{scope}
\begin{scope}[xshift=0cm, yshift=-4cm]
\draw[vector] (1,1) -- (-1,0);
\draw[scalar] (1,-1) -- (-1,0);
\draw[scalar] (1,-1) -- (1,1);
\draw[scalar] (-1,0) -- (-3,0);
\draw[vector] (3,1.3) -- (1,1);
\draw[vector] (3,-1.3) -- (1,-1);
\draw (-1,0) node [vertex] {};
\draw (1,1) node  [vertex] {};
\draw (1,-1) node [vertex] {};
\draw (-2,2) node {\small $(f)$};
\end{scope}
\begin{scope}[xshift=7cm, yshift=-4cm]
\draw[vector] (1,1) -- (-1,0);
\draw[vector] (1,-1) -- (-1,0);
\draw[scalar] (1,-1) -- (1,1);
\draw[scalar] (-1,0) -- (-3,0);
\draw[vector] (3,1.3) -- (1,1);
\draw[vector] (3,-1.3) -- (1,-1);
\draw (-1,0) node [vertex] {};
\draw (1,1) node  [vertex] {};
\draw (1,-1) node [vertex] {};
\draw (-2,2) node {\small $(g)$};
\end{scope}
\begin{scope}[xshift=14cm, yshift=-4cm]
\draw[scalar] (1,1) -- (-1,0);
\draw[scalar] (1,-1) -- (-1,0);
\draw[vector] (1,-1) -- (1,1);
\draw[scalar] (-1,0) -- (-3,0);
\draw[vector] (3,1.3) -- (1,1);
\draw[vector] (3,-1.3) -- (1,-1);
\draw (-1,0) node [vertex] {};
\draw (1,1) node  [vertex] {};
\draw (1,-1) node [vertex] {};
\draw (-2,2) node {\small $(h)$};
\end{scope}
\begin{scope}[xshift=-7cm, yshift=-8cm]
\draw[scalar] (1,1) -- (-1,0);
\draw[scalar] (1,-1) -- (-1,0);
\draw[scalar] (1,-1) -- (1,1);
\draw[scalar] (-1,0) -- (-3,0);
\draw[vector] (3,1.3) -- (1,1);
\draw[vector] (3,-1.3) -- (1,-1);
\draw (-1,0) node [vertex] {};
\draw (1,1) node  [vertex] {};
\draw (1,-1) node [vertex] {};
\draw (-2,2) node {\small $(i)$};
\end{scope}
\begin{scope}[xshift=0cm, yshift=-8cm]
\draw[scalar] (-1,0) -- (-3,0);
\draw[vector] (3,1.3) -- (1,0);
\draw[vector] (3,-1.3) -- (1,0);
\draw (-1,0) node  [vertex] {};
\draw (1,0) node  [vertex] {};
\draw[scalar] (1,0) arc (0:360:1);
\draw (-2,2) node {\small $(j)$};
\end{scope}
\begin{scope}[xshift=7cm, yshift=-8cm]
\draw[scalar] (-1,0) -- (-3,0);
\draw[vector] (3,1.3) -- (0,1);
\draw[vector] (3,-1.3) -- (0,-1);
\draw (0,1) node  [vertex] {};
\draw (0,-1) node  [vertex] {};
\draw[ghost] (1,0) arc (0:360:1);
\draw (-2,2) node {\small $(k)$};
\end{scope}
\end{tikzpicture}
\caption{\label{fig:AllDiagrams} 
One-loop Feynman diagrams for the process $h\to\ga\ga$. Diagram $(a)$ contains the fermion loops, while diagrams $(b)\,$--$\,(k)$ show the contributions from the gauge sector in a general $R_\xi$ gauge. Solid lines represent fermion mass eigenstates, wavy lines vector-boson mass eigenstates $W_\mu^{\pm(n)}$, dashed lines scalar mass eigenstates $\varphi_W^{\pm(n)}$, and dotted lines ghost mass eigenstates $c_W^{\pm(n)}$. The ghost masses and profiles are the same as for the $W$ bosons and their KK excitations \cite{Archer}.}
\end{center}
\end{figure}

The one-loop Feynman diagrams contributing to the $h\to\ga\ga$ decay amplitude are shown in Figure~\ref{fig:AllDiagrams} for a general $R_\xi$ gauge. In the subsequent section we will demonstrate that the full amplitude is gauge invariant. In the unitary gauge only the diagrams $(a)\,$--$\,(c)$ contribute. In order to present our results, we find it convenient to parametrize the $h\to\gamma\gamma$ amplitude, including the contributions from SM particles, by means of two Wilson coefficients $C_{1\gamma}$ and $C_{5\gamma}$ define via
\begin{equation}\label{eqn:effamp}
   \mathcal{A}(h\to\ga\ga) 
   = C_{1\gamma}\,\frac{\al}{6\pi v}\,\langle\ga\ga| F_{\mu\nu} F^{\mu\nu} |0\rangle 
    - C_{5\gamma}\,\frac{\al}{4\pi v}\,\langle\ga\ga| F_{\mu\nu} \tilde F^{\mu\nu} 
    |0\rangle \,,
\end{equation} 
where $\tilde F^{\mu\nu}=-\frac{1}{2}\e^{\mu\nu\al\be} F_{\al\be}$ with $\e^{0123}=-1$. Each Wilson coefficient can be written as a sum of three terms,
\begin{equation}
   C_i = C_i^W + C_i^q + C_i^l \,, 
\end{equation}
where in a general gauge $C_i^W$ includes the bosonic contributions from gauge bosons, scalar bosons, and ghosts. The calculation of these bosonic contributions is the main subject of this work. The fermionic loop contributions due to virtual quarks and leptons shown in diagram $(a)$ can be readily deduced from expressions derived in \cite{Malm:2013jia}. They will be summarized in Section~\ref{subsec:fermions}. 

In our analysis we will also discuss the case of a very narrow Higgs boson localized near the IR brane, where the Higgs profile $\delta^\eta(t-1)$ has a characteristic width $\eta$  subject to the condition $\mkk\ll v/\eta\ll\Lambda_{\rm TeV}$. In principle, such a scenario gives rise to a tower of {\em physical\/} scalar particles $\phi_W^{\pm(n)}$, which in some sense are the KK excitations of the charged components of the Higgs doublet. As discussed in detail in \cite{Archer}, these fields are defined in terms of a gauge-invariant superposition of $W_\phi^\pm$ and $\varphi^\pm$. It has been shown in the same reference that the effect of these heavy scalar particles on the $h\to\ga\ga$ amplitude is
\begin{equation}\label{C1phi}
   C_{1\gamma}^\phi = \frac{1}{8} \sum_{n=1}^\infty
    \frac{v g_{h\phi\phi}^{(n,n)}}{\big(m_n^\phi\big)^2}\,A_\phi(\tau_n^\phi) \,,
   \qquad C_{5\gamma}^\phi = 0 \,,
\end{equation}
where $\tau_n^\phi=4(m_n^\phi)^2/m_h^2$, and the function
\begin{equation}\label{scalarloop}
   A_\phi(\tau) = 3\tau\,\big[ \tau f(\tau) - 1 \big] \,, \quad
   \mbox{with} \quad
   f(\tau) = \arctan^2\frac{1}{\sqrt{\tau-1}} \,,
\end{equation}
approaches~1 for $\tau\to\infty$. In the limit of a very narrow Higgs profile the couplings $g_{h\phi\phi}^{(n,n)}$ scale like $1/\eta$, while the masses of the heavy scalar particles scale like $\mkk/\eta$. It follows that $C_{1\gamma}^\phi={\cal O}(\eta)$, and hence this contribution decouples in the limit $\eta\to 0$, as expected. We will therefore not consider the corresponding Feynman diagrams in our analysis.

\subsection{Fermionic contributions to the Wilson coefficients}
\label{subsec:fermions}

The one-loop contributions to the $h\to\ga\ga$ amplitude due to the exchange of virtual quarks and leptons can be derived in a straightforward way from analogous results for the quark contributions to the $gg\to h$ amplitude, which were studied in \cite{Casagrande:2010si,Azatov:2010pf,Azatov:2011qy,Goertz:2011hj,Carena:2012fk,Malm:2013jia}. Here we will use expressions derived in our previous work \cite{Malm:2013jia}, where a variety of RS models were considered using the same 5D approach employed in the present work. All that is necessary is to include appropriate factor of color and electric charges. The exact result can be written in the form
\begin{equation}\label{ourbeauty}
\begin{aligned}
   C_{1\gamma}^q 
   &= 3 N_c \sum_{f=u,d}\,Q_q^2\,\int_0^1\!dx \int_0^{1-x}\!dy\,(1-4xy)
    \left[\, T_+^q(-xy m_h^2) - T_+^q(\Lambda_{\rm TeV}^2) \right] , \\
   C_{5\gamma}^q 
   &= 2 N_c \sum_{f=u,d}\,Q_q^2\,\int_0^1\!dx \int_0^{1-x}\!dy
    \left[\, T_-^q(-xy m_h^2) - T_-^q(\Lambda_{\rm TeV}^2) \right] ,
\end{aligned}
\end{equation}
where $Q_u=2/3$ and $Q_d=-1/3$ are the electric charges of the quarks, and $N_c=3$ is the number of colors. The functions $T_\pm^q(-p^2)$ are defined in terms of linear combinations of overlap integrals of the Higgs-boson profile with the chirality-odd components of the 5D fermion propagator (see eq.~(15) in \cite{Malm:2013jia} for more details). An analogous expression, with $N_c$ replaced by~1 and $Q_q$ replaced by $Q_e=-1$ holds for the charged-lepton contribution. These exact results can be simplified by neglecting some terms of order $v^4/\mkk^4$ and chirally-suppressed $\ord(v^2/\mkk^2)$ terms, which is an excellent approximation numerically. This leads to the explicit expressions \cite{Malm:2013jia}
\begin{align}\label{eqn:cga}
   C_{1\gamma}^q 
   &\approx \left[ 1 - \frac{v^2}{3\mkk^2}\,{\rm Re}\,
    \frac{(\bm{Y}_u \bm{Y}_u^\dagger \bm{Y}_u )_{33}}{(\bm{Y}_u)_{33}} \right] 
    N_c\,Q_u^2\,A_q(\tau_t) + N_c\,Q_d^2\,A_q(\tau_b) 
    + \sum_{q=u,d} N_c\,Q_q^2\,\,{\rm Re}\,\tr\,g(\Xb_q) \,, \nonumber\\
   C_{5\gamma}^q
   &\approx - \frac{v^2}{3\mkk^2}\,{\rm Im} \left[
    \frac{(\bm{Y}_u \bm{Y}_u^\dagger \bm{Y}_u )_{33}}{(\bm{Y}_u)_{33}} \right] 
    N_c\,Q_u^2\,B_q(\tau_t) + \sum_{q=u,d} N_c\,Q_q^2\,\,{\rm Im}\,\tr\,g(\Xb_q) \,, 
\end{align}
and
\begin{equation}\label{leptonterm}
   C_{1\gamma}^l + i C_{5\gamma}^l \approx Q_e^2\,\,\tr\,g(\Xb_e) \,, 
\end{equation}
where the contributions from the SM fermions and the KK excitations can now readily be identified. The loop functions are given by (with $\tau_i=4m_i^2/m_h^2$)
\begin{equation}
   A_q(\tau) = \frac{3\tau}{2}\,\big[ 1 + (1-\tau) f(\tau) \big] \,, \qquad
   B_q(\tau) = \tau f(\tau) \,.
\end{equation} 
They both approach~1 for $\tau\to\infty$. For values $\tau<1$ the function $f(\tau)$ in (\ref{scalarloop}) must be analytically continued, with $\tau\to\tau-i0$. The quantities
\begin{equation}\label{Xfdef}
   \Xb_f = \frac{v}{\sqrt{2}\mkk}\,\sqrt{\bm{Y}_f\bm{Y}_f^\dagger} \,; \quad
   f = u,d,e
\end{equation}
are defined in terms of the dimensionless 5D Yukawa matrices of the RS model, whose entries are assumed to be random complex numbers of order~1. The observed hierarchies in the spectrum of fermion masses and mixing angles are generated when these anarchic Yukawa matrices are combined with the values of the fermion profiles near the IR brane, which are exponentially small for all light fermions \cite{Grossman:1999ra,Gherghetta:2000qt,Huber:2000ie}. Note that with the hermitian matrices $\Xb_f$ the traces over matrix-valued functions $g(\Xb_f)$ are real, so that $C_{5\gamma}^l=0$ and the only contribution to the coefficient $C_{5\gamma}^q$ arises from the top-quark contribution given by the first term on the right-hand side of (\ref{eqn:cga}). The precise form of the function $g(\bm{X}_f)$ depends on the details of the localization of the scalar sector on or near the IR brane. For the two scenarios with a brane-localized Higgs and a narrow bulk Higgs, as defined in Section~\ref{sec:pre}, one finds \cite{Carena:2012fk,Malm:2013jia}
\begin{equation}\label{eqn:gfunc}
\begin{aligned}
   g(\bm{X}_f) \big|_{\rm brane\;Higgs} &= - \frac{\bm{X}_f\tanh\bm{X}_f}{\cosh2\bm{X}_f}
    = - \bm{X}_f^2 + \ord\bigg( \frac{v^4}{\mkk^4} \bigg) \,, \\ 
   g(\bm{X}_f) \big|_{\rm narrow\;bulk\;Higgs} &= \bm{X}_f\tanh\bm{X}_f
    = \bm{X}_f^2 + \ord\bigg( \frac{v^4}{\mkk^4} \bigg) \,,
\end{aligned}
\end{equation}
so that the effect of the KK tower is approximately equal but of opposite sign in the two cases. The difference is due to a ``resonance effect'' in the narrow bulk-Higgs scenario, where very heavy KK modes with masses of order the inverse Higgs width $\Delta_h=v/\eta$ give an unsuppressed contribution to the loop amplitude \cite{Carena:2012fk,Delaunay:2012cz}. At a technical level, the difference arises from the subtraction term at large Euclidean momentum in (\ref{ourbeauty}), which is relevant for the function $T_+^f(-p^2)$ only. For a brane-localized Higgs, this function approaches a plateau at large momenta, such that $T_+^f(\Lambda_{\rm TeV}^2)=\tr\,\bm{X}_f\tanh2\bm{X}_f$. For a narrow bulk Higgs, on the other hand, the function $T_+^f(p_E^2)$ vanishes like $1/p_E$ in the region of large Euclidean momenta $p_E^2=-p^2\gg (v/\eta)^2$, and hence $T_+^f(\Lambda_{\rm TeV}^2)$ can be set to zero. In \cite{Malm:2013jia} we have also considered a variant of the brane-Higgs scenario with two different Yukawa matrices $\bm{Y}_f^C$ and $\bm{Y}_f^S$ for the $Z_2$-even and $Z_2$-odd fermion fields. In this so-called type-II brane-Higgs model the matrices $\bm{X}_f$ are no longer hermitian, but to leading order
\begin{equation}\label{typeII}
   g(\bm{Y}_f^C,\bm{Y}_f^S) \big|_{\rm type-II\;brane\;Higgs} 
   = - \frac{v^2}{2\mkk^2}\,\bm{Y}_f^C \bm{Y}_f^{C\dagger}
    + \ord\bigg( \frac{v^4}{\mkk^4} \bigg) 
\end{equation}
is still a hermitian matrix. The type-II brane-Higgs scenario is thus rather similar to the original brane-Higgs model with identical Yukawa matrices $\bm{Y}_f^C=\bm{Y}_f^S=\bm{Y}_f$. Numerically, we find that the main difference is a slightly larger spread of the distribution of scatter points when one scans over the parameter space of the model. In our phenomenological analysis in Section~\ref{sec:pheno}, we will therefore restrict ourselves to a study of the two cases shown in (\ref{eqn:gfunc}). 

\subsection{Gauge invariance of the amplitude} 
\label{sec:gauge}

In the SM, a recent paper \cite{Marciano:2011gm} has thoroughly discussed the $\xi$ independence of the $h\to\ga\ga$ amplitude in dimensional regularization and has shown that the calculation can be performed consistently in the unitary gauge $\xi\to\infty$. In the case of the RS model, it is convenient to first work in the KK-decomposed theory, where 4D Feynman propagators have the same structure as in the SM. The Feynman rules required to evaluate the one-loop diagrams shown in Figure~\ref{fig:AllDiagrams} are summarized in Appendix~\ref{app:FR}. From these rules, it follows that:
\begin{itemize}
\item
All vertices involving one or two external photons but no Higgs boson are diagonal in KK number after one integrates over the extra-dimensional coordinate of the vertex with measure $\int_{-\pi}^\pi d\phi=(2\pi/L)\int_\epsilon^1\!dt/t$. The Feynman rules for these vertices have the same form as in the SM after one identifies the 4D electromagnetic coupling as $e=e_5/\sqrt{2\pi r}$ \cite{Davoudiasl:1999tf,Pomarol:1999ad}. For the mass-dependent vertex connecting a photon to $W_\mu^{\pm(n)}\varphi_W^{\mp(n)}$, one must replace $m_W\to m_n^W$. 
\item
As a result, all one-loop diagrams contributing to the $h\to\ga\ga$ amplitude involve a single KK particle in the loop. Hence, only KK-diagonal Higgs couplings are required in the calculation.
\item
All KK-diagonal Higgs couplings have the same structure as in the SM but come with an overall prefactor
\begin{equation}\label{prefac}
   \frac{v}{2}\,\frac{g_5^2}{2\pi r}\,2\pi \left[ \chi_n^W(1) \right]^2
   = \frac{2\tilde m_W^2}{v}\,2\pi \left[ \chi_n^W(1) \right]^2 ,
\end{equation}
which replaces the corresponding factor $g m_W=2m_W^2/v_{\rm SM}$ in the SM. In addition, for each scalar boson $\varphi_W^{\pm(n)}$ a factor $1/m_n^W$ appears, which replaces $1/m_W$ in the corresponding SM Feynman rule for vertices involving the Goldstone bosons $\varphi^\pm$. 
\end{itemize}
It follows from these observations that, diagram by diagram and in a general $R_\xi$ gauge, the bosonic loop contributions obtained in the RS model resemble those of the SM up to trivial substitutions, such that
\begin{equation}\label{eqn:super}
   \mathcal{A}_{\rm RS}^W(h\to\ga\ga) 
   = \frac{\tilde m_W^2}{v}\,\sum_{n=0}^\infty\,2\pi \left[ \chi_n^W(1) \right]^2
    \left[ \frac{v_{\rm SM}}{m_W^2}\,\mathcal{A}_\SM^W(h\to\ga\ga) \right]_{m_W\to m_n^W} .
\end{equation}

The statement made in the first bullet point above requires some comments. For vertices involving a photon and a pair of vector bosons, fermions or ghosts, the statement that the vertex is diagonal in KK number is a direct consequence of the flatness of the photon profile, enforced by $U(1)_{\rm EM}$ gauge invariance, and the orthogonality of the relevant vector-boson and fermion profiles. However, the situation is different for vertices involving the scalar bosons $\varphi_W^{\pm(n)}$, which receive contributions from the 5D fields $W_\phi^\pm$ and $\varphi^\pm$, see (\ref{eqn:KKdecomM}) and (\ref{varphidecomp}). In this case the vertex becomes diagonal only after one adds up these two contributions. Consider, as an example, the vertex
\begin{center}
\begin{tikzpicture}[scale=0.35]
\node (origin) at (0,0) {};
\begin{scope}[xshift=-16cm, yshift=0cm]
\draw[vector] (15,0) -- (17, -2)node[right]{$A_\mu^{(0)}$};
\draw[vector](15,0)--(17,2)node[right]{$A^{(0)}_\nu$};
\draw[scalar](13,2) node[left] { $\varphi^{\pm(n)}$}--(15,0);
\draw[scalar](13,-2) node[left] { $\varphi^{\mp(m)}$}--(15,0);
\end{scope}
\end{tikzpicture}
\end{center}
needed for diagram $(j)$ in Figure~\ref{fig:AllDiagrams}. After integrating over the coordinate of this vertex, we obtain the Feynman rule 
\begin{equation}
   2ie^2\eta_{\mu\nu} \left[ \frac{\mkk^2}{m_m^W\,m_n^W}\,
    \frac{2\pi}{L} \int_\epsilon^1\!\frac{dt}{t} \left[ \partial_t\chi_m^W(t) \right]
    \left[ \partial_t\chi_n^W(t) \right] 
    + \frac{\tilde m_W^2}{m_m^W\,m_n^W}\,2\pi\,\chi_m^W(1)\,\chi_n^W(1) \right] ,
\end{equation}
where the first contribution originates from the $W_\phi W_\phi A_\mu A^\mu$ term contained in the Yang-Mills action for the $W$-boson fields using the KK decomposition (\ref{eqn:KKdecomM}), while the second contribution arises from the $\varphi^+\varphi^- A_\mu A^\mu$ term contained in the kinetic term for the Higgs doublet using the KK decomposition (\ref{varphidecomp}). We now integrate by parts in the first term use the equations of motion \cite{Davoudiasl:1999tf,Pomarol:1999ad}
\begin{equation}\label{chinDEQ}
   - t\,\partial_t\,\frac{1}{t}\,\partial_t\chi_n^W(t) 
   = \frac{\left( m_n^W\right)^2}{\mkk^2}\,\chi_n^W(t)
\end{equation}
for the gauge-boson profiles along with the boundary conditions
\begin{equation}\label{chinBCs}
   \partial_t\chi_n^W(t) \big|_{t=\epsilon} = 0 \,, \qquad
   \partial_t\chi_n^W(t) \big|_{t=1^-} = - \frac{L\tilde m_W^2}{\mkk^2}\,\chi_n^W(1)
\end{equation}
corresponding to a Neumann boundary condition on the UV brane and a mixed boundary condition on the IR brane. The notation $t=1^-$ means that the IR brane is approached from the left ($t<1$). Such a prescription is necessary because the derivative of the profile function is discontinuous on the IR brane. In this way, we obtain the Feynman rule
\begin{equation}
   2ie^2\eta_{\mu\nu} \left[ \frac{m_n^W}{m_m^W}\,
    \frac{2\pi}{L} \int_\epsilon^1\!\frac{dt}{t}\,\chi_m^W(t)\,\chi_n^W(t) \right] ,
\end{equation}
where the boundary term cancels the contribution arising from the $\varphi^+\varphi^- A_\mu A^\mu$ term. Using finally the orthonormality relation \cite{Casagrande:2008hr,Davoudiasl:1999tf,Pomarol:1999ad}
\begin{equation}\label{orthonorm}
   \frac{2\pi}{L} \int_\epsilon^1\!\frac{dt}{t}\,\chi_m^W(t)\,\chi_n^W(t) = \delta_{mn}
\end{equation}
for the gauge-boson profiles, we obtain the SM expression $2ie^2\eta_{\mu\nu}\,\delta_{mn}$ for the vertex.

Let us now return to the general result (\ref{eqn:super}) and explore its consequences. Obviously, this relation implies that for each single KK mode the $h\to\ga\ga$ amplitude in the RS model is gauge invariant provided the amplitude is gauge invariant in the SM. Since, as we will demonstrate below, the sum over KK modes is convergent, it follows that gauge invariance is maintained also in the 5D theory. We recall that to show gauge invariance in the SM one divides the $W$-boson propagator in $R_\xi$ gauge into two parts,
\begin{equation}
   \frac{i}{p^2-m_W^2} \left[ \frac{(1-\xi)\,p^\mu p^\nu}{p^2-\xi m_W^2}
    - \eta^{\mu\nu} \right] 
   = \frac{i}{p^2-m_W^2} \left( \frac{p^\mu p^\nu}{m_W^2} - \eta^{\mu\nu} \right) 
    - \frac{i}{p^2-\xi m_W^2}\,\frac{p^\mu p^\nu}{m_W^2} \,,
\end{equation}
where the first part coincides with the propagator in unitary gauge and the second part has the same structure as the scalar-boson and ghost propagators. It has been shown in \cite{Marciano:2011gm} that, after adding up all diagrams, many intricate cancellations occur, and at the end only the diagrams $(b)$ and $(c)$ in Figure~\ref{fig:AllDiagrams} with the $W$-boson propagators in unitary gauge, as well as the fermion loop contributions shown in diagram $(a)$, remain. We have repeated this analysis and checked these cancellations by explicit calculation, thereby confirming that it is justified to use unitary gauge also in the RS model.

\subsection{5D analysis of the bosonic loop contributions to $\bm{h\to\ga\ga}$}
\label{sec:3.3}

We now repeat the calculation of the bosonic loop contributions to the $h\to\ga\ga$ amplitude using a 5D approach. Based on the findings of the previous section we adopt unitary gauge and consider only the contributions of diagrams $(b)$ and $(c)$ in Figure~\ref{fig:AllDiagrams}. We employ the mixed momentum-position representation of the 5D gauge-boson propagator $D^\xi_{W,\mu\nu}(t,t';p)$ \cite{Randall:2001gb,Puchwein:2003jq,Contino:2004vy,Carena:2004zn,Csaki:2010aj}, in which the extra-dimensional coordinate is kept in position space. This is very convenient from a technical point of view, but it is also physically well motivated, as by AdS/CFT correspondence the position along the extra dimension defines the natural mass scale of the model \cite{Randall:2001gb,Gherghetta:2010cj}. It is well known that even in the SM the loop-momentum integral must be regularized dimensionally in order to preserve gauge invariance. We will thus introduce a dimensional regulator $d=4-2\hat\epsilon$ on the loop integral in intermediate steps. This regulator can be removed at the end of the calculation. We also regularize the Higgs profile by replacing the $\delta$-function profile of the brane-localized Higgs field by a smooth function $\delta^\eta(t-1)$ of width $\eta\ll 1$. Such a regularization is important in the calculation of the fermionic loop contributions to the $gg\to h$ and $h\to\ga\ga$ amplitudes. However, we will find that in the calculation of the bosonic loop contributions the limit $\eta\to 0$ can be taken without encountering any ambiguities.

Diagrams $(b)$ and $(c)$ give rise to the amplitude
\begin{equation}\label{eqn:5Damp}
\begin{aligned}
   i &\mathcal{A}(h\to\ga\ga) 
   = - \frac{2\tilde m_W^2}{v}\,2\pi e^2\,\e^*_\mu(k_1)\,\e^*_\nu(k_2)\,\eta^{\al\be}
    \int\frac{d^dp}{(2\pi)^d} \int_\e^1\!dt\,\delta^\eta(t-1)\, 
    \frac{2\pi}{L} \int_\e^1\!\frac{dt_1}{t_1} \\ 
   &\times \bigg[ \frac{2\pi}{L} \int_\e^1\!\frac{dt_2}{t_2}\,
    2V^{\gamma\mu\lambda\rho\nu\delta}\, 
    D^{\xi\to\infty}_{W,\alpha\gamma}(t,t_1,p+k_1)\,
    D^{\xi\to\infty}_{W,\lambda\rho}(t_1,t_2,p)\,
    D^{\xi\to\infty}_{W,\delta\beta}(t_2,t,p-k_2) \\ 
   &\hspace{7mm}\mbox{}+ \left( 2\eta^{\gamma\delta} \eta^{\mu\nu}
    - \eta^{\delta\nu} \eta^{\gamma\mu} - \eta^{\nu\gamma} \eta^{\mu\delta} \right)
    D^{\xi\to\infty}_{W,\al\gamma}(t,t_1,p+k_1)\, 
    D^{\xi\to\infty}_{W,\be\delta}(t_1,t,p-k_2) \bigg] \,,
\end{aligned}
\end{equation}
where $V^{\gamma\mu\lambda\rho\nu\delta}=V^{\gamma\mu\lambda}(p+k_1,-k_1,-p)\,V^{\rho\nu\delta}(p,-k_2,-p+k_2)$ arises from the product of two triple gauge-boson vertices, with
$V^{\mu\nu\rho}(k,p,q)=\eta^{\mu\nu} (k-p)^\rho + \eta^{\nu\rho} (p-q)^\mu + \eta^{\rho\mu} (q-k)^\nu$. Our goal is to rewrite this result as a Feynman parameter integral over a {\em single\/} 5D gauge-boson propagator, which should be possible since in the KK-decomposed theory only a single KK mode propagates in the loops. In order to simplify the answer, we decompose the 5D propagator as
\begin{equation}\label{eqn:5DPropKK}
   D^\xi_{W,\mu\nu}(t,t';p) 
   = B_W(t,t';-p^2-i0) \left( \eta_{\mu\nu}-\frac{p_\mu p_\nu}{p^2} \right) 
    + B_W(t,t';-p^2/\xi-i0)\,\frac{p_\mu p_\nu}{p^2} \,,
\end{equation} 
and use the KK decomposition
\begin{equation}\label{level42}
   B_W(t,t';-p^2-i0) 
   = \sum_{n=0}^\infty\,\frac{\chi^W_n(t)\,\chi^W_n(t')}{\left(m^W_n\right)^2-p^2-i0}
\end{equation}
in an intermediate step. The use of the KK representation is merely a mathematical trick, similar to the use of Feynman parameters in conventional loop calculations. It is justified because all expressions are finite and the KK sum converges. At the end of the calculation we obtain an expression without any reference to KK modes.

Due to the mode-diagonality of the vertices involving a photon, we can perform the integration over $t_1$ and $t_2$ using the orthonormality relation (\ref{orthonorm}) for the gauge-boson profiles. Working out the Dirac algebra and making use of Passarino-Veltman reductions, we can reduce the answer to a simple Feynman parameter integral. After the contributions from the various diagrams have been combined, the dimensional regulator $\hat\epsilon$ can be set to~0. We find
\begin{equation}\label{eq23}
   C_{1\gamma}^W = - 3\pi\,\tilde m_W^2 \int_\e^1\!dt\,\delta^\eta(t-1)  
    \sum_{n=0}^\infty \left[\chi_n^W(t) \right]^2  
    \left[ \frac{1}{\left(m_n^W\right)^2} + 6\int_0^1\!dx\int_0^{1-x}\!dy\,
    \frac{1-2xy}{\left(m_n^W\right)^2-xy m_h^2 - i0} \right] ,
\end{equation}
and $C_{5\gamma}^W=0$. It is now a simple exercise to recast the answer in terms of the 5D propagator function $B_W(t,t';-p^2)$ defined in (\ref{eqn:5DPropKK}). We obtain
\begin{equation}\label{eqn:C1W}
   C_{1\gamma}^W = - 3\pi\tilde m_W^2 \bigg[ T_W(0) 
    + 6\int_0^1\!dx\int_0^{1-x}\!dy\,(1-2xy)\,T_W(-xy m_h^2) \bigg] \,,
\end{equation}
where $T_W(-p^2)$ denotes the overlap integral of the Higgs profile with the transverse part of the 5D $W$-boson propagator evaluated at $t=t'$,
\begin{equation}\label{TWdef}
   T_W(-p^2) = \int_\e^1\!dt\,\delta^\eta(t-1)\,B_W(t,t;-p^2-i0) 
   = B_W(1,1;-p^2-i0) + {\cal O}(\eta) \,.
\end{equation}
For the case of a brane-localized scalar sector, an explicit expression for the function $B_W$ will be derived in the following section. The more general case of a narrow bulk Higgs will be briefly considered at the end of Section~\ref{sec:AnalZeroMRS} and will be described in more detail elsewhere. Such an analysis demonstrates that the above integral exhibits a smooth behavior in the limit of small $\eta$, so that the last identity holds and the regulator on the Higgs profile can be taken to zero without encountering any ambiguities.

Relation~\eqref{eqn:C1W} is one of the main results of this work. It shows the exact result for the Wilson coefficient $C_{1\gamma}$ in dependence of overlap integrals of the Higgs profile and the 5D gauge-boson propagator. With the help of the findings in \cite{Archer}, it can be shown that this relation also holds for an arbitrary bulk-Higgs profile $\chi_h(t)$, provided one uses the corresponding 5D gauge-boson propagator in the bulk-Higgs model. Then the regularized $\delta$-function in (\ref{TWdef}) must be replaced by
\begin{equation}\label{bulkreplace}
   \delta^\eta(t-1)\to\frac{2\pi}{Lt}\,\frac{v(t)}{v}\,\chi_h(t) 
   = 2(1+\beta)\,t^{1+2\beta} + \dots \,,
\end{equation}
where $v(t)$ is the profile of the Higgs vev, and we use the conventions of \cite{Malm:2013jia}. Note, however, that in this case it is necessary to also include the contribution (\ref{C1phi}) due to the physical scalar excitations of the bulk Higgs field. The last equation in (\ref{bulkreplace}) holds (up to very small corrections) in the particular bulk-Higgs model analyzed in \cite{Cacciapaglia:2006mz,Archer:2012qa}. The parameter $\beta>0$ is related to the 5D mass parameter of the bulk scalar field. In the region where $\beta\gg 1$, the function on the right-hand side indeed approaches a regularized $\delta$-distribution, with a characteristic width given by $\eta=1/(2\beta)$. 

Note that relation (\ref{eq23}) results after integrating a Feynman loop integrand of the type $1/[p_E^2+(m_n^W)^2-xy m_h^2]^3$ over $d^4p_E$ (after the Wick rotation), which corresponds to the integral over the second derivative $\partial_{p_E^2}^2 T_W(p_E^2-xym_h^2)$. In order for this integral to exist, we need to require that both $T_W(p_E^2)$ and $p_E\,\partial_{p_E} T_W(p_E^2)$ vanish for very large Euclidean momenta. We will show in the following section that this is indeed the case.

\subsection{Calculation of the 5D gauge-boson propagator}

The calculation of the propagator function $B_W$ in (\ref{eqn:5DPropKK}) in the RS model with a brane-localized Higgs field is a straightforward exercise and has been performed, for instance, in \cite{Randall:2001gb,Csaki:2002gy,Goertz:2011gk}. This function is the solution to the differential equation (with $p^2\equiv p^2+i0$)
\begin{equation}\label{DEG}
   \left( t\,\partial_t\,\frac{1}{t}\,\partial_t + \hat p^2 \right) B_W(t,t';-p^2)
   = - \frac{Lt}{2\pi\mkk^2}\,\delta(t-t') \,; \qquad
   \hat p^2 = \frac{p^2}{\mkk^2} \,,
\end{equation}
subject to the boundary conditions \cite{Casagrande:2008hr}
\begin{equation}\label{eqn:PropBC}
\begin{aligned}
   \left. \pa_t B_W(t,t';-p^2) \right|_{t=\e} &= 0 \,, \\
   \left. \pa_t B_W(t,t';-p^2) \right|_{t=1^-} 
   &= b_1 B_W({1},t';-p^2) \,, \qquad b_1 = - \frac{L\tilde m_W^2}{\mkk^2} \,.
\end{aligned}
\end{equation}
Note the close similarity with the corresponding equations for the gauge-boson profiles $\chi_n^W(t)$ in (\ref{chinDEQ}) and (\ref{chinBCs}). Integrating the differential equation (\ref{DEG}) over an infinitesimal interval around $t=t'$, we derive the jump condition 
\begin{equation}
   \left. \pa_t B_W(t,t';-p^2) \right|_{t=t'-0}^{~t'+0} = - \frac{Lt'}{2\pi\mkk^2} \,.
\end{equation}
The propagator itself is continuous at $t=t'$. In the region of time-like momenta ($p^2\ge 0$), the general solution can be written in the form
\begin{equation}\label{eqn:BbraneMin}
   B_W(t,t';-p^2) 
   = \frac{Ltt'}{4\mkk^2}\,
    \frac{\left[ \hat p D_{10}(t_>,1) - b_1 D_{11}(t_>,1) \right] D_{10}(t_<,\e)}%
         {\hat p D_{00}(1,\e) - b_1 D_{10}(1,\e)} \,,
\end{equation}
where 
\begin{equation}\label{eqn:Dij}
   D_{ij}(t,t') = J_i(\hat p t)\,Y_j(\hat p t') - Y_i(\hat p t)\,J_j(\hat p t') \,.\end{equation}
For space-like momenta, we find instead (with $p_E^2=-p^2>0$ and $\hat p_E^2=p_E^2/\mkk^2$)
\begin{equation}
   B_W(t,t';p_E^2) 
   = \frac{Ltt'}{2\pi\mkk^2}\,
    \frac{\left[ \pEh D_{10}(t_>,1) + b_1 D_{11}(t_>,1) \right] D_{10}(t_<,\e)}%
         {\pEh D_{00}(1,\e) - b_1 D_{10}(1,\e)} \,,
\end{equation}
with
\begin{equation}
   D_{ij}(t,t') = I_i(\pEh t)\,K_j(\pEh t') - (-1)^{i+j} K_i(\pEh t)\,I_j(\pEh t') \,.
\end{equation}

For our result (\ref{TWdef}) we need the propagator in the time-like region, evaluated at $t=t'=1$. Using the general solution in \eqref{eqn:BbraneMin}, we obtain (with $\hat p\equiv p/\mkk+i0$)
\begin{equation}\label{eqn:IW2}
   T_W(-p^2) 
   = \frac{1}{2\pi\tilde m_W^2} \left[ 1  + \frac{\hat p\mkk^2}{L\tilde m_W^2}   
    \frac{J_0(\hat p)\,Y_0(\hat p\e) - Y_0(\hat p)\,J_0(\hat p\e)}%
         {J_1(\hat p)\,Y_0(\hat p\e) - Y_1(\hat p)\,J_0(\hat p\e)} \right]^{-1}
   \equiv \frac{1}{2\pi\tilde m_W^2}\,\hat T_W(-p^2) \,,
\end{equation}
which is exact to all orders in $v^2/\mkk^2$.\footnote{The result can be simplified using that $J_0(\hat p\e)=1+{\cal O}(\epsilon^2)$ and $Y_0(\hat p\e)=(2/\pi)(\gamma_E+\ln(\hat p/2)-L)+{\cal O}(\epsilon^2)$.} 
It follows from this expression that $\hat T_W(0)=1$. We have thus succeeded in deriving a closed analytic expression for the Wilson coefficient $C^W_{1\gamma}$ in \eqref{eqn:C1W}, valid for the minimal RS model with a Higgs sector localized on the IR brane. Note that we have kept the quantity $\tilde m_W$, which is the leading-order contribution to the mass of the physical $W$ boson, in the prefactor above, since it will cancel against a corresponding factor in the definition of the Wilson coefficient \eqref{eqn:C1W}. Indeed, our final result for this coefficient takes the form
\begin{equation}
   C_{1\gamma}^W = - \frac32\,\bigg[ 1 + 6\int_0^1\!dx\int_0^{1-x}\!dy\,(1-2xy)\, 
    \hat T_W(-xy m_h^2) \bigg] \,.
\end{equation}

Before proceeding, we briefly study the behavior of the propagator function in the region of large space-like momenta. For large Euclidean momenta $p_E\gg\mkk$, this function approaches an inverse power-law behavior given by
\begin{equation}
   T_W(p_E^2) = \frac{L}{2\pi\mkk}\,\frac{1}{p_E} + {\cal O}(p_E^{-2}) \,.
\end{equation}
It follows that both $T_W(p_E^2)$ and $p_E\,\partial_{p_E} T_W(p_E^2)$ vanish for large Euclidean momenta $p_E^2=-p^2\to\infty$, and hence the conditions required for the validity of our relation (\ref{eqn:C1W}) are indeed satisfied. 

\subsection{Analysis of the zero-mode and KK contributions} 
\label{sec:AnalZeroMRS}

Our exact expression for the overlap integral $T_W(-p^2)$ in (\ref{eqn:IW2}) contains the contribution of the zero mode -- the standard $W$ boson with its modified coupling to the Higgs field -- as well as the infinite tower of KK excitations. It is instructive to isolate the contribution from the zero mode and the KK tower explicitly. To this end, we expand the exact formula in powers of $v^2/\mkk^2$, using that we need this function for values $p^2={\cal O}(m_h^2)$ much smaller than the KK scale $\mkk^2$. We find
\begin{equation}\label{eqn:TImin}
   \hat T_W(-p^2) = \frac{m_W^2}{m_W^2 - p^2 - i0}
    \left[ 1 - \frac{m_W^2}{2\mkk^2} \left( \frac{L}{c_\vartheta^2} - 1 + \frac{1}{2L} \right) \right] 
    + \frac{m_W^2}{2\mkk^2} \left( \frac{L}{c_\vartheta^2} - 1 + \frac{1}{2L} \right) 
    + {\cal O}\bigg( \frac{v^4}{\mkk^4} \bigg) \,, 
\end{equation}
where $c_\vartheta=1$ in the minimal RS model. In Section~\ref{sec:propcust} we will show that the same result holds in the custodial RS model, where however the parameter $c_\vartheta=1/\sqrt2$ takes a different value. In the above result we have replaced the parameter $\tilde m_W$ by the physical $W$-boson mass $m_W$ using relation (\ref{Wmassrela}), which was derived in \cite{Casagrande:2008hr} by solving the eigenvalue equation for the $W$-boson profiles and extracting the lowest eigenvalue. In Appendix~\ref{sec:massren}, we present an alternative approach, where the above relation is derived with the help of our expressions for the 5D gauge-boson propagator. 

Based on the formulas above, we can perform the integration over the Feynman parameters in \eqref{eqn:C1W} and find the Wilson coefficient
\begin{equation}\label{eqn:C1Wmin}
   C_{1\gamma}^W = - \frac{21}{4}\,\big[ \kappa_W A_W(\tau_W) + \nu_W \big] 
    + {\cal O}\bigg( \frac{v^4}{\mkk^4} \bigg) \,, \qquad
   C_{5\gamma}^W = 0 \,,
\end{equation}
where $\tau_W=4m_W^2/m_h^2$, and the function
\begin{equation}
   A_W(\tau) = \frac17\,\big[ 2 + 3\tau + 3\tau(2-\tau)\,f(\tau) \big]
\end{equation}
with $f(\tau)$ from (\ref{scalarloop}) approaches~1 for $\tau\to\infty$. The first contribution to $C_{1\gamma}$ arises from the standard $W$ boson, whose coupling to the Higgs boson is modified, compared with the SM, by a factor $\kappa_W$ times $v_{\rm SM}/v$. The last factor is accounted for by using the Higgs vev in the RS model in the definition of the effective operators in (\ref{eqn:effamp}). The term $\nu_W$ in (\ref{eqn:C1Wmin}) is due to the KK excitations. Explicitly, we obtain
\begin{equation}\label{eqn:cWMRS}
   \kappa_W = 1- \frac{m_W^2}{2\mkk^2} \left( \frac{L}{c_\vartheta^2} - 1 + \frac{1}{2L} \right) , \qquad
   \nu_W = \frac{m_W^2}{2\mkk^2} \left( \frac{L}{c_\vartheta^2} - 1 + \frac{1}{2L} \right) .
\end{equation}
Note that at this order $\nu_W=(1-\kappa_W)$, such that the RS corrections to $C_{1\gamma}$ in (\ref{eqn:C1Wmin}) would cancel in the limit $\tau_W\to\infty$. This simple relation is however not preserved in higher orders. Our result for $C_{1\gamma}$ agrees with a corresponding expression derived in \cite{Bouchart:2009vq}. Notice also that the value of $\kappa_W$ is consistent with relation (\ref{prefac}), which gives $\kappa_W=\frac{\tilde m_W^2}{m_W^2}\,2\pi[\chi_0^W(1)]^2$. 

We close this section by returning briefly to the case of a narrow bulk-Higgs model, in which the scalar sector is localized not on but near the IR brane. As a concrete model, we adopt the scenario discussed in \cite{Cacciapaglia:2006mz,Archer:2012qa,Malm:2013jia} featuring a bulk scalar field, which acquires a vev due to a mexican-hat potential localized on the IR brane. As discussed earlier, relation (\ref{TWdef}) still holds in this model provided one makes the replacement (\ref{bulkreplace}) and calculated the gauge-boson propagator in the background of a bulk-Higgs field. It is the solution to the differential equation
\begin{equation}
   \left( t\,\partial_t\,\frac{1}{t}\,\partial_t + \hat p^2 
    - \frac{2\pi\tilde m_W^2}{\mkk^2}\,\frac{v^2(t)}{v^2} \right) B_W(t,t';-p^2)
   = - \frac{Lt}{2\pi\mkk^2}\,\delta(t-t') \,,
\end{equation}
with Neumann boundary conditions $\partial_t B_W(t,t',-p^2)=0$ for $t=\epsilon, 1$. We will present the details of such an analysis elsewhere. Here it suffices to note that in the limit where $\eta=1/(2\beta)\ll 1$, one obtains 
\begin{equation}
   \kappa_W^{\rm bulk\,Higgs} 
    = \kappa_W + \frac{3L m_W^2}{2\mkk^2}\,\eta + \ord(\eta^2) \,, \qquad
   \nu_W^{\rm bulk\,Higgs} 
    = \nu_W - \frac{L m_W^2}{\mkk^2}\,\eta + \ord(\eta^2)
\end{equation}
instead of (\ref{eqn:C1Wmin}). This demonstrates that the result for the bosonic loop contributions to the $h\to\ga\ga$ amplitude interpolates smoothly from the narrow bulk-Higgs scenario into a scenario with a brane-localized scalar sector.

\section{Extension to the RS model with custodial symmetry}
\label{sec:custodial}

We will now present the generalization of the above results to the RS model with custodial protection, which has been proposed to mitigate the large corrections to electroweak precision observables, so that the lightest KK particles are in reach for the direct detection at the LHC \cite{Carena:2006bn,Cacciapaglia:2006gp,Contino:2006qr,Carena:2007ua}. We will consider an RS model based on the enlarged bulk symmetry $SU(3)_c\times SU(2)_L\times SU(2)_R\times U(1)_X\times P_{LR}$, whose $SU(2)$ subgroups are broken on the IR brane via the symmetry-breaking pattern $SU(2)_L\times SU(2)_R\to SU(2)_V$. This symmetry breaking is accomplished by means of the Higgs field transforming as a bi-doublet under the two $SU(2)$ gauge groups. In component notation, it is given by 
\begin{equation}\label{bidoublet}
   \Phi(x) = \frac{1}{\sqrt{2}} \begin{pmatrix}
    v + h(x) - i\varphi^3(x) & -i\sqrt{2}\,\varphi^+(x) \\
    -i\sqrt{2}\,\varphi^-(x) & v + h(x) + i\varphi^3(x) \end{pmatrix} ,
\end{equation}
where $\varphi^i$ are real scalar fields, $\varphi^\pm=(\varphi^1\mp i\varphi^2)/\sqrt{2}$, and $v$ denotes the Higgs vev in the custodial RS model. The resulting $SU(2)_V$ supplies the custodial symmetry and protects the $T$ parameter \cite{Agashe:2003zs,Csaki:2003zu}. The $P_{LR}$ symmetry prevents the left-handed $Zb\bar b$ coupling \cite{Agashe:2006at} and its flavor-changing counterparts \cite{Blanke:2008zb} from receiving too large corrections. On the UV brane, the symmetry breaking $SU(2)_R\times U(1)_X\to U(1)_Y$ generates the SM gauge group. This is achieved by an interplay between UV and IR boundary conditions. Thorough discussions of this model containing many technical details can be found in \cite{Casagrande:2010si,Albrecht:2009xr}, and we will adopt the notations of the first reference throughout this analysis.

\subsection{Quark contributions to the Wilson coefficients}
\label{sec:resultcust}

As a consequence of the discrete $P_{LR}$ symmetry, the left-handed bottom quark needs to be embedded in an $SU(2)_L\times SU(2)_R$ bi-doublet with isospin quantum numbers $T_L^3=-T_R^3=-1/2$. This assignment fixes the quantum numbers of the remaining quark fields uniquely. In particular, the right-handed down-type quarks have to be embedded in an $SU(2)_R$ triplet in order to obtain a $U(1)_X$-invariant Yukawa coupling. One arrives at the following multiplet structure for the quark fields with even $Z_2$ parity:
\begin{equation}\label{quarkmultiplets}
\begin{aligned}
   Q_L &= \left( \begin{array}{cc}
     {u_L^{(+)}}_{\frac 23} & {\lambda_L^{(-)}}_{\frac 53} \\
     {d_L^{(+)}}_{-\frac 13} & {u_L^{\prime\,(-)}}_{\frac 23}
    \end{array} \right)_{\!\!\frac 23} , \qquad 
    u_R^c = \left( {u_R^{c\,(+)}}_{\frac 23} \right)_{\frac 23} , \\
    {\cal T}_R = {\cal T}_{1R}\oplus{\cal T}_{2R}
    &= \left( \begin{array}{c}
      {\Lambda_R^{\prime\,(-)}}_{\frac 53} \\
      {U_R^{\prime\,(-)}}_{\frac 23} \\
      {D_R^{\prime\,(-)}}_{-\frac 13}
     \end{array} \right)_{\!\!\frac 23}
     \oplus \left( {D_R^{(+)}}_{-\frac 13} \quad {U_R^{(-)}}_{\frac 23} \quad 
      {\Lambda_R^{(-)}}_{\frac 53} \right)_{\frac 23} .
\end{aligned}
\end{equation}
The field $Q_L$ transforms as $(\bm{2},\bm{2})$ under $SU(2)_L\times SU(2)_R$, while ${\cal T}_R$ transforms as $(\bm{3},\bm{1})\oplus(\bm{1},\bm{3})$. The fields with odd $Z_2$ parity have the opposite chirality. Their profiles are related to those of the $Z_2$-even fields by the field equations. The inner and outer subscripts on the various fields denote their $U(1)_{\rm EM}$ and $U(1)_X$ charges, respectively, which are connected through the relations $Y=-T_R^3+Q_X$ and $Q=T_L^3+Y$. The superscripts on the fields specify the type of boundary conditions they obey on the UV boundary. Fields with superscript $(+)$ obey the usual mixed boundary conditions allowing for a light zero mode, meaning that we impose a Dirichlet boundary condition on the profile functions of the corresponding $Z_2$-odd fields. These zero modes correspond to the SM quarks. Fields with superscripts $(-)$ correspond to heavy, exotic fermions with no counterparts in the SM. For these states, the Dirichlet boundary condition is imposed on the $Z_2$-even fields, so as to avoid the presence of a zero mode. The UV boundary conditions for the fields of opposite $Z_2$ parity are of mixed type and follow from the field equations. 

Note that we chose the same $SU(2)_L\times SU(2)_R$ quantum numbers for all three quark generations, which is necessary to consistently incorporate quark mixing in the anarchic approach to flavor in warped extra dimensions. Altogether, there are fifteen different quark states in the up sector and nine in the down sector (for three generations). The boundary conditions give rise to three light modes in each sector, which are identified with the SM quarks. These are accompanied by KK towers consisting of groups of fifteen and nine modes of similar masses in the up and down sectors, respectively. In addition, there is a KK tower of exotic fermion states with electric charge $Q_\lambda=5/3$, which exhibits nine excitations in each KK level. 

The fermionic loop contributions to the $h\to\ga\ga$ amplitude in the custodial RS model can be parameterized in terms of the same $3\times 3$ Yukawa matrices appearing in the minimal model, however with different coefficients that reflect the embeddings of the various fermion species under the enlarged bulk gauge group. The generalizations of relations (\ref{eqn:cga}) for the quark contributions have been worked out in \cite{Malm:2013jia}. They are
\begin{equation}\label{eqn:Cqcust}
\begin{aligned}
   C_{1\gamma}^q 
   &\approx \left[ 1 - \frac{2v^2}{3\mkk^2}\,{\rm Re}\,
    \frac{(\bm{Y}_u \bm{Y}_u^\dagger \bm{Y}_u )_{33}}{(\bm{Y}_u)_{33}} \right] 
    N_c\,Q_u^2\,A_q(\tau_t) + N_c\,Q_d^2\,A_q(\tau_b) \\
   &\quad\mbox{}+ N_c\,Q_u^2\,\,{\rm Re}\,\tr\,g(\sqrt{2}\Xb_u) 
    + N_c \left( Q_u^2 + Q_d^2 + Q_\lambda^2 \right) {\rm Re}\,\tr\,g(\sqrt{2}\Xb_d) \,, \\
   C_{5\gamma}^q
   &\approx - \frac{2v^2}{3\mkk^2}\,{\rm Im} \left[
    \frac{(\bm{Y}_u \bm{Y}_u^\dagger \bm{Y}_u )_{33}}{(\bm{Y}_u)_{33}} \right] 
    N_c\,Q_u^2\,B_q(\tau_t) \\
   &\quad\mbox{}+ N_c\,Q_u^2\,\,{\rm Im}\,\tr\,g(\sqrt{2}\Xb_u) 
    + N_c \left( Q_u^2 + Q_d^2 + Q_\lambda^2 \right) {\rm Im}\,\tr\,g(\sqrt{2}\Xb_d) \,.
\end{aligned}
\end{equation}
For various RS models with a brane-localized scalar sector or a narrow bulk-Higgs sector, the explicit forms of the function $g(\bm{X}_f)$ have been given in \eqref{eqn:gfunc} and \eqref{typeII}. Recall that the Taylor expansion of these functions starts with $\bm{X}_f^2$, and thus the factors of $\sqrt{2}$ arising in the quark contributions in the custodial model approximately double the contribution arising in the minimal model. Combined with the large electric charge of the $\lambda$-type quarks, one finds that due to the higher multiplicity of KK quark states the contribution in the custodial RS model is much larger than in the minimal model \cite{Goertz:2011hj,Malm:2013jia}, by approximately a factor 68/5. 

\subsection{Charged-lepton contributions to the Wilson coefficients}

The result for the loop contributions to the $h\to\ga\ga$ amplitude involving charged leptons depends on the way in which the lepton fields are embedded into the extended gauge symmetry of the custodial RS model. As a first possibility, we consider a model in which the lepton multiplets are chosen in analogy to the quark multiplets in (\ref{quarkmultiplets}). This choice was adopted in \cite{Albrecht:2009xr}. In component notation, the corresponding fields are
\begin{equation}\label{lepton1}
\begin{aligned}
   \xi_{1L} &= \left( \begin{array}{cc}
     {\nu_L^{(+)}}_0 & {\psi_L^{(-)}}_1 \\
     {e_L^{(+)}}_{-1} & {\nu_L^{\prime\,(-)}}_0
    \end{array} \right)_{\!\! 0} , \qquad 
   \xi_{2R} = \left( {\nu_R^{c\,(+)}}_0 \right)_0 , \\
   \xi_{3R} = {\cal T}_{3R}\oplus{\cal T}_{4R}
   &= \left( \begin{array}{c}
     {\Psi_R^{\prime\,(-)}}_1 \\
     {N_R^{\prime\,(-)}}_0 \\
     {E_R^{\prime\,(-)}}_{-1}
    \end{array} \right)_{\!\! 0}
    \oplus \left( {E_R^{(+)}}_{-1} \quad {N_R^{(-)}}_0 \quad 
     {\Psi_R^{(-)}}_1 \right)_0 .
\end{aligned}
\end{equation}
There are fifteen different lepton states in the neutrino sector and nine in the charged-lepton sector. The boundary conditions give rise to three light modes in each sector, which are identified with the SM neutrinos and charged leptons. These are accompanied by KK towers consisting of groups of fifteen and nine modes in the two sectors, respectively. In addition, there is a KK tower of exotic lepton states with electric charge $Q_\psi=+1$, which exhibits nine excitations in each KK level. The gauge-invariant Yukawa interactions for these fields are constructed in complete analogy with the quark Yukawa interactions \cite{Casagrande:2010si,Albrecht:2009xr}. They can be expressed in terms of two dimensionless $3\times 3$ Yukawa matrices $\bm{Y}_\nu$ and $\bm{Y}_e$, which we assume to have an anarchic structure. When dressed with the fermion profiles on the IR brane, these matrices give masses to the SM leptons. The resulting contributions to the Wilson coefficients have the same structure as in (\ref{eqn:Cqcust}), except that there are no zero-mode contributions (they are proportional to $m_l^2/m_h^2$ and thus can be neglected) and that we must replace $\bm{Y}_u\to\bm{Y}_\nu$, $\bm{Y}_d\to \bm{Y}_e$, $N_c\to 1$, and $Q_u\to Q_\nu=0$, $Q_d\to Q_e=-1$, $Q_\lambda\to Q_\psi=+1$. We thus obtain
\begin{equation}
   C_{1\gamma}^l + iC_{5\gamma}^l
   \approx \left( Q_e^2 + Q_\psi^2 \right) \tr\,g(\sqrt{2}\Xb_e) \,,
\end{equation}
with $\bm{X}_e$ as defined in (\ref{Xfdef}). It follows that the leptonic contribution in the custodial RS model is approximately 4 times larger than in the minimal model. 

As a second possibility, we consider a model with a more minimal embedding of the leptons into the extended gauge group. The simplest assignment is to put the left-handed neutrino and electron into an $SU(2)_L$ doublet (as in the SM) and the right-handed electron along with a new, exotic neutral particle $N_R$ into an $SU(2)_R$ doublet. The lepton fields with even $Z_2$ parity are then chosen as
\begin{equation}\label{lepton2}
   L_L = \begin{pmatrix} {\nu_L^{(+)}}_0 \\ {e_L^{(+)}}_{-1} \end{pmatrix}_{-\frac12} ,
    \qquad 
   L_R^c = \begin{pmatrix} {e_R^{c(+)}}_{-1} \\ {N_R^{(-)}}_{0}
    \end{pmatrix}_{-\frac12} ,
\end{equation}
and they transform as $(\bm{2},\bm{1})$ and $(\bm{1},\bm{2})$, respectively. The choice of the boundary conditions is such that the zero modes correspond to the light leptons of the SM, without a right-handed neutrino. The gauge-invariant Yukawa interaction that can be built using these fields is
\begin{equation}\label{eqn:Yuklep}
   {\cal L}_Y 
   = \frac{v}{\sqrt2} \int_{-\pi}^\pi\!d\phi\,\delta(|\phi|-\pi)\,e^{-3\sigma(\phi)}\,
    \frac{2}{k}\,\big( Y_e \big)_{ij} \big( 
    \bar L_L^i\Phi\,\varepsilon L_R^{c\,j} + \bar L_R^i\Phi\,\varepsilon L_L^{c\,j}
    \big) + \mbox{h.c.} \,,
\end{equation} 
where $\varepsilon=i\sigma^2$. Upon electroweak symmetry breaking this generates a mass term for the zero modes of the charged leptons. The SM neutrinos remain massless at this order. Their masses can be explained by means of higher-dimensional operators. The only additional lepton field is the right-handed neutrino, which is charged under $SU(2)_R$ but electrically neutral, so that it does not affect the $h\to\ga\ga$ decay amplitude. The lepton contribution is therefore the same as in the minimal version of the RS model, namely $C_{1\gamma}^l+iC_{5\gamma}^l\approx Q_e^2\,\tx{Tr}\,g(\Xb_e)$ as in (\ref{leptonterm}). 

\subsection{The bosonic sector}

In order to derive the Feynman rules and the 5D gauge-boson propagator it is inevitable to understand the bosonic sector of the custodial RS model, whose 5D action reads
\begin{equation}
   S_{\rm gauge} = \int d^4x\,\frac{2\pi r}{L} \int_{\e}^1\!\frac{dt}{t} 
    \left( \La_{L,R,X} + \La_{\rm Higgs} + \La_{\rm GF} \right) .
\end{equation}
Since it is of no significance for our discussion, we refrain from presenting the gauge-fixing term $\La_{\rm GF}$, whose explicit form can be found in \cite{Casagrande:2010si}. The gauge-kinetic terms read
\begin{equation}
   \La_{L,R,X} = \frac{\sqrt{G}}{r}\,G^{KM} G^{LN}
    \left( - \frac14\,L_{KL}^a L_{MN}^a - \frac14\,R_{KL}^a R_{MN}^a 
    - \frac14\,X_{KL} X_{MN} \right) ,
\end{equation}
where $G^{MN}$ denotes the 5D metric with determinant $G=r^2 e^{-8\sigma(\phi)}$, and where a sum over the gauge-group indices $a=1,2,3$ is implied. We choose the 4-components of the gauge fields to be even under the $Z_2$ parity, while the fifth components are chosen to be odd, in order to derive at a low-energy spectrum that is compatible with observation. As in the previous section, it does not make any difference if we consider a narrow-bulk or a brane-localized Higgs sector, and we thus focus on the scenario in which the Higgs Lagrangian is localized on the IR brane. The Higgs transforms as a bi-doublet under $SU(2)_L\times SU(2)_R$ and is neutral with respect to $U(1)_X$, see (\ref{bidoublet}). In order to show how the symmetry breaking $SU(2)_L\times SU(2)_R\to SU(2)_V$ is accomplished, we use the covariant derivative
\begin{equation}
   D_\mu\Phi = \partial_\mu\Phi - ig_{L,5} L_\mu^a T^a_L\,\Phi + ig_{R,5} \Phi R_\mu^a T^a_R \,,
\end{equation}
where $g_{L,5}$ and $g_{R,5}$ are the 5D gauge couplings associated with $SU(2)_{L,R}$, and $T_{L,R}^a=\sigma^a/2$ are the corresponding generators. In order to evaluate the kinetic term for the scalar bi-doublet, it is convenient to rotate the gauge bosons $L_\mu^a$ and $R_\mu^a$ into a new basis of fields $\tilde A_\mu^a$ and $V_\mu^a$, such that \cite{Burdman:2008gm}
\begin{equation}\label{basistrafo}
   \begin{pmatrix} \tilde A_M^a \\ V_M^a \end{pmatrix} 
   = \begin{pmatrix} \cos\vartheta && -\sin\vartheta \\ 
                     \sin\vartheta && \phantom{-}\cos\vartheta \end{pmatrix}
    \begin{pmatrix} L_M^a \\ R_M^a \end{pmatrix} 
   \equiv \bm{R}_\vartheta \begin{pmatrix} L_M^a \\ R_M^a \end{pmatrix} ,
\end{equation}
where
\begin{equation}
   \cos\vartheta = \frac{g_{L,5}}{\sqrt{g_{L,5}^2 + g_{R,5}^2}} \,,	\qquad
   \sin\vartheta = \frac{g_{R,5}}{\sqrt{g_{L,5}^2 + g_{R,5}^2}} \,.
\end{equation}
The $P_{LR}$ symmetry, which is imposed to protect the left-handed $Z\bar b b$ couplings from receiving large corrections \cite{Agashe:2006at}, enforces that $g_{L,5}=g_{R,5}$, and hence $\cos\vartheta=\sin\vartheta=1/\sqrt2$. In our discussion in this section we will however keep the value of $\vartheta$ as a free parameter. The Higgs vev $\langle\Phi\rangle=(v/\sqrt2)\,\bm{1}$ then generates a mass term $M_{\tilde A}^2=v^2(g_{L,5}^2+g_{R,5}^2)/4$ for the fields $\tilde A_\mu^a$, while the fields $V_\mu^a$ remain massless. We can also read off the coupling to the Higgs boson, once we replace $v^2$ by $(v+h)^2$. Note that only the fields $\tilde A_\mu^a$ couple to the Higgs boson $h$. This will become important for the derivation of the propagator. The further symmetry breaking via boundary conditions is not relevant for our discussion, and we again refer to \cite{Casagrande:2010si} for details. Notice that relation (\ref{basistrafo}) represents the connection between the UV basis fields (right), which obey Dirichlet boundary conditions on the UV brane, and the IR basis fields (left), which obey Dirichlet boundary conditions on the IR brane. 

We now focus on the charged sector and define the gauge-boson fields
\begin{equation}
   \vec W_M^\pm \equiv \begin{pmatrix} \tilde A_M^\pm \\ V_M^\pm \end{pmatrix} 
   = \bm{R}_\vartheta \begin{pmatrix} L_M^\pm \\ R_M^\pm \end{pmatrix} 
   = \frac{\bm{R}_\vartheta}{\sqrt2}
    \begin{pmatrix} L_M^1 \mp i L_M^2 \\ R_M^1 \mp i R_M^2 \end{pmatrix} ,
\end{equation}
whose KK decomposition can be written in a form analogous to (\ref{eqn:KKdecomM}), such that \cite{Casagrande:2010si}
\begin{equation}\label{eqn:KKdecomC}
\begin{aligned}
   \vec W_\mu^\pm(x,t) &= \frac{\bm{R}_\vartheta}{\sqrt r} \sum_{n=0}^\infty
    \vec{\chi}_n^{\,W}(t)\,W_\mu^{\pm(n)}(x) \,, \\
   \vec W_\phi^\pm(x,t) &= -\frac{\bm{R}_\vartheta}{\sqrt r}\,
    \frac{L}{\pi} \sum_{n=0}^\infty\,\frac{1}{m_n^W}\,
    t\,\partial_t\vec{\chi}_n^{\,W}(t)\,\varphi_W^{\pm(n)}(x) \,.
\end{aligned}
\end{equation}
The orthonormality relation for the gauge-boson profiles reads
\begin{equation}\label{eqn:oncust}
   \frac{2\pi}{L} \int_{\epsilon}^1\!\frac{dt}{t}\,\vec{\chi}_n^{\,W}(t)^T\,
    \vec{\chi}_m^{\,W}(t) = \delta_{nm} \,.
\end{equation} 

The profiles $\vec{\chi}_n^{\,W}(t)$ are $Z_2$-even functions on the orbifold. Their upper (lower) components are ``untwisted'' (``twisted'') functions. Untwisted even functions obey Neumann boundary conditions on the UV brane, allowing for light zero modes. Twisted even functions obey Dirichlet boundary conditions on the UV brane and are thus not smooth at this orbifold fixed point. The upper (lower) components of the rotated profiles $\bm{R}_\vartheta \vec{\chi}_n^{\,W}(t)$ obey mixed (Neumann) boundary conditions on the IR brane, such that
\begin{equation}\label{IRbcs}
   \bm{R}_\vartheta\,\partial_t\vec{\chi}_n^{\,W}(t) \big|_{t=1^-}
   = - \frac{L\tilde m_W^2}{c_\vartheta^2\mkk^2}\,\bm{P}_+ 
    \bm{R}_\vartheta\,\vec{\chi}_n^{\,W}(1) \,; \qquad
   \tilde m_W = \frac{g_{L,5}}{\sqrt{2\pi r}}\,\frac{v}{2} \,,
\end{equation}
which generalizes (\ref{chinBCs}). Here $\bm{P}_+=\mbox{diag}(1,0)$ is a projector on the upper component, and from now on we use the abbreviations $c_\vartheta\equiv\cos\vartheta$ and $s_\vartheta\equiv\sin\vartheta$. As in the minimal RS model, the parameter $\tilde m_W$ is the leading contribution to the $W$-boson mass in an expansion in powers of $v^2/\mkk^2$, see (\ref{Wmassrela}).

It is now straightforward to deduce the Feynman rules in the custodial model from the ones in the minimal model compiled in Appendix~\ref{app:FR}. Using \eqref{eqn:oncust}, we can convince ourselves that the $W_M^\pm$ couplings to the photon are not changed at all. This statement is independent of the basis, since the rotation matrix $\bm{R}_\vartheta$ drops out in the orthonormalization condition. In contrast, as mentioned earlier the Higgs only couples to the IR basis fields $\tilde A_\mu^\pm$ with a strength proportional to $(g_{L,5}^2+g_{R,5}^2)$. This can be taken into account with the help of the projection operator $\bm{P}_+$ rotated into the IR basis and accompanied by a factor $1/c_\vartheta^2$. It follows that, compared with the SM, all KK-diagonal Higgs couplings in the custodial RS model come with a prefactor
\begin{equation}
   \frac{2\tilde m_W^2}{c_\vartheta^2\,v}\,2\pi\,
    \vec\chi_n^{\,W}(1)^T\,\bm{R}_\vartheta^T\,\bm{P}_+\,
    \bm{R}_\vartheta\,\vec\chi_n^{\,W}(1) 
   \equiv \frac{2\tilde m_W^2}{c_\vartheta^2\,v}\,2\pi\,
    \vec\chi_n^{\,W}(1)^T\,\bm{D}_\vartheta\,\vec\chi_n^{\,W}(1) \,,
\end{equation}
which replaces the corresponding factor (\ref{prefac}) in the minimal model. Here we have introduced
\begin{equation}\label{eqn:DW}
   \bm{D}_\vartheta = \bm{R}_\vartheta^T\,\bm{P}_+\,\bm{R}_\vartheta 
   = \begin{pmatrix} c_\vartheta^2 & -s_\vartheta c_\vartheta \\
                     -s_\vartheta c_\vartheta & ~s_\vartheta^2 \end{pmatrix} .
\end{equation}

In analogy with (\ref{level42}), we now define the propagator function 
\begin{equation}\label{eqn:propKK}
   \bm{B}_W^{\rm UV}(t,t';-p^2-i0) = \sum_{n=0}^\infty\,
    \frac{\vec{\chi}_n^{\,W}(t)\,\vec{\chi}_n^{\,W}(t')^T}{\left(m^W_n\right)^2-p^2-i0}
\end{equation}
in terms of gauge-boson profiles in the UV basis. An explicit expression for this function will be derived in the next section. In analogy with expression (\ref{eqn:super}) valid in minimal RS model, we find that the $h\to\ga\ga$ amplitude in the custodial RS model can be written as 
\begin{equation}\label{eqn:ampcust}
   \mathcal{A}_{\rm cust.\,RS}^W(h\to\ga\ga) 
   = \frac{\tilde m_W^2}{c_\vartheta^2\,v}\,\sum_{n=0}^\infty\,2\pi\,
    \vec\chi_n^W(1)^T \bm{D}_\vartheta\,\vec\chi_n^W(1)
    \left[ \frac{v_{\rm SM}}{m_W^2}\,\mathcal{A}_\SM^W(h\to\ga\ga) \right]_{m_W\to m_n^W} .
\end{equation}
It follows that expression (\ref{eqn:C1W}) for the Wilson coefficient $C_{1\gamma}^W$ derived in Section~\ref{sec:3.3} remains valid, provided we replace the quantity $T_W(-p^2)$ defined in (\ref{TWdef}) with
\begin{equation}\label{TWdefcusto}
   T_W(-p^2) = \mbox{Tr} \left[ 
    \frac{\bm{D}_\vartheta}{c_\vartheta^2}\,\bm{B}_W^{\rm UV}(1,1;-p^2-i0) \right] .
\end{equation}

\subsection{Calculation of the 5D gauge-boson propagator}
\label{sec:propcust}

We now derive the exact expression for the 5D gauge-boson propagator in the RS model with custodial symmetry, which to the best of our knowledge has not been done before. The differential equation for the propagator function $\bm{B}_{\rm UV}$ is the same as in the minimal model, see (\ref{DEG}). However, the boundary conditions are modified to \cite{Casagrande:2010si} 
\begin{equation}\label{eqn:UVBounds}
\begin{aligned}
   \left( \bm{P}_+\,\partial_t + \bm{P}_- \right) \bm{B}_W^{\rm UV}(t,t';-p^2)
    \big|_{t=\epsilon} 
   &= 0 \,, \\
   \left( \partial_t - b_1 \bm{D}_\vartheta \right) \bm{B}_W^{\rm UV}(t,t';-p^2)
    \big|_{t=1^-}
   &= 0 \,; \qquad b_1 = - \frac{L\tilde m_W^2}{c_\vartheta^2\mkk^2} \,.
\end{aligned}
\end{equation} 
The first equation follows from the boundary conditions for the UV fields $L_M^\pm$ and $R_M^\pm$, where we have defined $\bm{P}_-=\mbox{diag}(0,1)$. The second equation is a direct consequence of (\ref{IRbcs}). We find that, in the region of time-like momenta ($p^2>0$), the general solution for the propagator function reads
\begin{equation}\label{eqn:BUV}
\begin{aligned}
   \bm{B}_W^{\rm UV}(t,t';-p^2) 
   &= \frac{Ltt'}{4\mkk^2}\,
    \frac{1}{\left[ \hat p D_{00}(1,\e) - b_1 D_{10}(1,\e) \right] D_{01}(1,\e)
             - b_1 \frac{4s_\vartheta^2}{\pi^2\hat p^2\e}} \\
   &\times \bigg\{ \left[ 
    \big[ \hat p D_{10}(t_>,1) - b_1 D_{11}(t_>,1) \big] D_{01}(1,\e) 
    - b_1 \frac{2s_\vartheta^2}{\pi\hat p}\,D_{11}(t_>,\e) \right] 
    D_{10}(t_<,\e)\,\bm{P}_+ \\
   &\quad\mbox{}+ \left[ 
    \big[ \hat p D_{00}(1,\e) - b_1 D_{10}(1,\e) \big] D_{10}(t_>,1) 
    + b_1 \frac{2s_\vartheta^2}{\pi\hat p}\,D_{10}(t_>,\e) \right] 
    D_{11}(t_<,\e)\,\bm{P}_- \\
   &\quad\mbox{}- b_1 \frac{2s_\vartheta c_\vartheta}{\pi\hat p} 
    \big[ D_{10}(t,\e)\,D_{11}(t',\e)\,\bm{P}_{12} 
    + D_{11}(t,\e)\,D_{10}(t',\e)\,\bm{P}_{21} \big] \bigg\} \,,
\end{aligned}
\end{equation}
where the functions $D^\pm_{ij}(t,t')$ have been defined in \eqref{eqn:Dij}, and we have introduced the $2\times 2$ matrices $\bm{P}_{12}$ and $\bm{P}_{21}$, which have an entry~1 at the corresponding position indicated by the subscripts and entries~0 otherwise. Note that up to irrelevant $\ord(\epsilon^2)$ terms we can replace $\hat p\e D_{n1}(t,\e)=-\frac{2}{\pi}\,J_n(\hat p t)$ for $n=0,1$. This gives rise to a simpler expression, in which the spurious $1/\epsilon$ term in the denominator is removed. In the limit $s_\vartheta\to 0$, we can identify the coefficient of $\bm{P}_+$ in (\ref{eqn:BUV}) with the result (\ref{eqn:BbraneMin}) obtained in the minimal RS model. Moreover, for the special case $p^2=0$ our result reduces to equation~(54) in \cite{Casagrande:2010si}. Our general results above valid for arbitrary momentum are however new.

It is now straightforward to calculate the quantity $T_W(-p^2)$ in (\ref{TWdefcusto}), which we need for the calculation of the Wilson coefficient $C_{1\gamma}$ in (\ref{eqn:C1W}). Expanding this answer in powers of $v^2/\mkk^2$ and for $p^2=\ord(m_h^2)$, we recover expression (\ref{eqn:TImin}). With respect to the minimal RS model, the only modification concerns the coefficient of the leading $L$-enhanced correction terms, which is enhanced by $1/c_\vartheta^2$. This affects both the contributions from the $W$ boson and the KK tower. In the custodial RS model with $P_{LR}$ symmetry, this enhancement factor is equal to~2. Note that with $c_\vartheta^2=1/2$ the expressions in (\ref{eqn:cWMRS}) are compatible with  corresponding results obtained in \cite{Casagrande:2010si}. In this reference the Wilson coefficient $C_{1\gamma}$ belonged to the operator $v h F_{\mu\nu} F^{\mu\nu}$ instead of the one in \eqref{eqn:effamp}, and hence $\kappa_W^{\rm Ref.~[19]}=\kappa_W/\kappa_v^2$.

\section{Phenomenological implications}
\label{sec:pheno}

We now present a numerical study of the Higgs decay into two photons in both the minimal and the custodial RS model, which can be directly compared to experimental data. As in our recent work on Higgs production \cite{Malm:2013jia}, we distinguish the two cases of a brane-localized and a narrow bulk-Higgs scenario. We consider the ratio of the measured $pp\to h\to\gamma\gamma$ cross section normalized to its SM value,
\begin{equation}\label{eqn:Rgaga}
   R_{\gamma\gamma} 
   = \frac{(\sigma\cdot{\rm Br)}(pp\to h\to\gamma\gamma)_{\rm RS}}%
          {(\sigma\cdot{\rm Br)}(pp\to h\to\gamma\gamma)_\SM} 
   = \frac{\big[ \big(|\kappa_g|^2 + |\kappa_{g5}|^2\big) f_{\rm GF} + \kappa_W^2 f_{\rm VBF} \big] 
           \big( |\kappa_{\gamma}|^2 + |\kappa_{\gamma5}|^2 \big)}%
          {\kappa_v^2\,\kappa_h} \,,
\end{equation}
where we have included the two main Higgs production channels via gluon fusion (GF) and vector-boson fusion (VBF), with probabilities of $f_{\rm GF}\approx 0.9$ and $f_{\rm VBF}\approx 0.1$ at the LHC with $\sqrt{s}=8$\,TeV \cite{Heinemeyer:2013tqa}. Other Higgs production channels, such as the associated production with a $t\bar t$ pair or a vector boson, can be neglected to a very good approximation. The quantities $\kappa_i$ and $\kappa_{i5}$ (with $i=g,\gamma$) parametrize the values of the relevant Wilson coefficients normalized to their SM values,
\begin{equation}
   \kappa_i = \frac{C_{1i}}{C_i^{\rm SM}} \,, \qquad  
   \kappa_{i5} = \frac{3}{2}\,\frac{C_{5i}}{C_i^{\rm SM}} \,.
\end{equation}
Explicit expressions for the Wilson coefficients $C_{1\gamma}$ and $C_{5\gamma}$ in the RS model have been derived in Sections~\ref{sec:analysis} and \ref{sec:custodial}. The corresponding SM value is $C_\gamma^{\rm SM}=N_c\big[ Q_u^2\,A_q(\tau_t)+Q_d^2\,A_q(\tau_b) \big]-\frac{21}{4} A_W(\tau_W)$. The RS effects on the gluon-fusion production process were studied in \cite{Malm:2013jia}. The values of the Wilson coefficients $C_{1g}$ and $C_{5g}$ can be obtained from $C_{1\gamma}$ and $C_{5\gamma}$ by replacing $Q_{u,d}\to 1$, $Q_e\to 0$, and $N_c\to 1$. In the SM we have $C_g^{\rm SM}=A_q(\tau_t)+A_q(\tau_b)$. Concerning the VBF production process, we note that using $\kappa_W$ as a correction factor in (\ref{eqn:Rgaga}) is only approximate but sufficient for our purposes \cite{phenopaper}. 

The parameter $\kappa_v$ in (\ref{eqn:Rgaga}) parameterizes the shift of the Higgs vev in the RS model relative to the SM and has been given in (\ref{kappav}). Finally, we take into account the RS corrections to the SM Higgs width $\Gamma^\SM_h=4.14$\,MeV (for $m_h=125.5$\,GeV) \cite{Denner:2011mq} by means of the parameter 
\begin{equation}\label{eqn:kh}
   \kappa_h = {\kappa_v^2}\,\frac{\Gamma_h^\tx{RS}}{\Gamma_h^\tx{SM}}
   \approx 0.57\,\kappa_b^2 + 0.22\,\kappa_W^2
    + 0.09\,\big( |\kappa_g|^2 + |\kappa_{g5}|^2 \big) + 0.12 \,, 
\end{equation}
where the corrections to the decays $h\to\tau^+\tau^-,\,c\bar c,\,ZZ^{(*)},\ga\ga,\dots$ have a numerically insignificant effect and therefore can be neglected (the combined branching ratio for these channels is 12\% in the SM). Neglecting some small chirally-suppressed terms, the correction to the Higgs coupling to a $b\bar b$ pair can be well approximated by \cite{Casagrande:2010si,Goertz:2011hj,Carena:2012fk}
\begin{equation}
   \kappa_b^{\rm min.\,RS} \approx  1 - \frac{v^2}{3\mkk^2}\,
    \frac{(\bm{Y}_d \bm{Y}_d^\dagger \bm{Y}_d )_{33}}{(\bm{Y}_d)_{33}} \,, \qquad
   \kappa_b^{\rm cust.\,RS} \approx  1 - \frac{2v^2}{3\mkk^2}\,
    \frac{(\bm{Y}_d \bm{Y}_d^\dagger \bm{Y}_d )_{33}}{(\bm{Y}_d)_{33}} \,.
\end{equation}

It is an important goal of future LHC and ILC analyses to determine as many of the effective Higgs couplings $\kappa_i$ as possible from a global fit to the data. A detailed discussion of the individual effective Higgs couplings to fermions and gauge bosons in the context of RS models will be presented in a future work \cite{phenopaper}. At present, however, the experimental groups have not yet presented a detailed, model-independent analysis of Higgs couplings \cite{Aad:2013wqa,CMSresult,Plehn:2012iz}, and we will thus focus on the ratio $R_{\ga\ga}$ in the present work. Note also that, in contrast to the observable $R_{\gamma\gamma}$, the quantities $\kappa_i$ and $\kappa_{i5}$ are not directly observable. The gluon-fusion rate is proportional to the sum of the absolute squares of $\kappa_g$ and $\kappa_{g5}$, and no observable sensitive to a different combination of these parameters is experimentally accessible. In the case of $h\to\ga\ga$ decay, it is in principle possible to access the CP-violating coefficient $\kappa_{\gamma 5}$ by studying the distribution of the two electron-positron pairs in events in which both photons undergo nuclear conversions \cite{Bishara:2013vya}, however this will be very challenging experimentally.

\begin{figure}
\begin{center}
\psfrag{x}[]{\small $M_{g^{(1)}}~{\rm [TeV]}$}
\psfrag{y}[b]{\small $R_{\gamma\gamma}$}
\psfrag{z}[]{\small \begin{tabular}{l} minimal RS model \\[-1mm] brane Higgs \end{tabular}}
\psfrag{w}[]{\small \begin{tabular}{l} minimal RS model \\[-1mm] bulk Higgs \end{tabular}}
\psfrag{a}[bl]{\footnotesize $~y_\star=0.5$}
\psfrag{b}[bl]{\footnotesize $~y_\star=1.5$}
\psfrag{c}[bl]{\footnotesize $~y_\star=3$}
\includegraphics[width=1.03\textwidth]{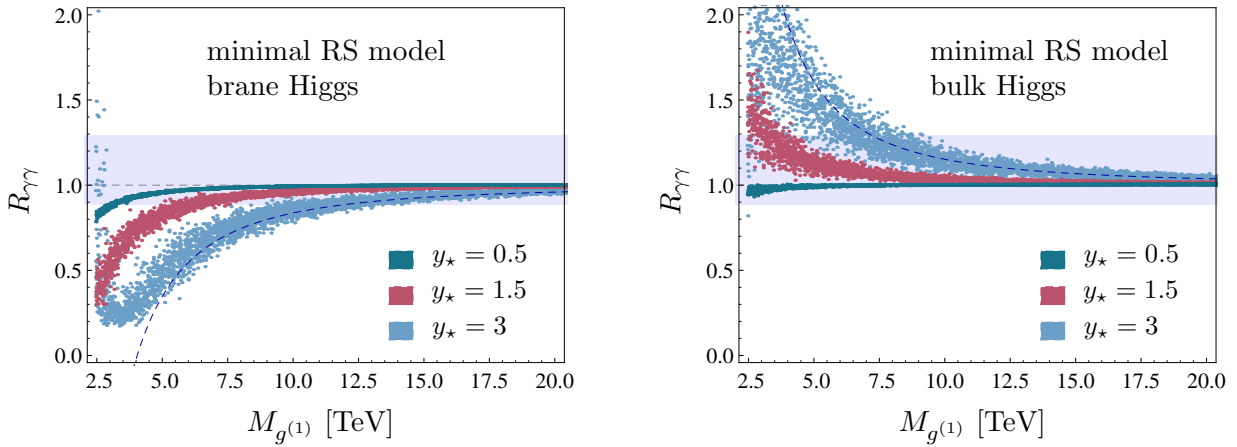}
\parbox{15.5cm}
{\caption{\label{fig:RggMRS}
Predictions for the ratio $R_{\gamma\gamma}$ as a function of the lightest KK gluon mass $M_{g^{(1)}}$ and for different values of the parameter $y_\star$ in the minimal RS model, for the cases of a brane-localized Higgs boson (left) and a narrow bulk-Higgs field (right). The dashed curves show the approximation (\ref{eqn:RgaMin}) for $y_\star=3$.}}
\end{center}
\end{figure}

Figure~\ref{fig:RggMRS} shows our predictions for $R_{\gamma\gamma}$ obtained in the minimal RS model with a brane-localized Higgs sector (left plot) and a narrow bulk-Higgs state (right plot). The new-physics effects arising in these scenarios scale with $1/\mkk^2$. We find it useful to convert the mass parameter $\mkk$ to the physical mass $M_{g^{(1)}}\approx 2.45\mkk$ of the lightest KK gluon (or KK photon) state, which is independent of the details of the localization of the scalar sector and of the choice of the electroweak gauge group in the bulk \cite{Davoudiasl:1999tf,Pomarol:1999ad}. Our numerical results also depend on the RS volume $L=\ln(M_{\rm Pl}/\Lambda_{\rm TeV})$ and the dimensionless 5D Yukawa matrices $\bm{Y}_u$, $\bm{Y}_d$ and $\bm{Y}_l$. Typical values for $L$ fall in the range $33\,\mbox{--}\,34$, corresponding to $\Lambda_{\rm TeV}\sim 20\,\mbox{--}\,50$\,TeV, and we take $L=33.5$ for concreteness. We work with anarchic Yukawa matrices, whose individual entries are taken to be random complex numbers subject to the condition that $0\le|(\bm{Y}_f)_{ij}|\le y_\star$. As a further constraint, we impose that these matrices correctly reproduce the Wolfenstein parameters $\rho$ and $\eta$ of the unitarity triangle (see \cite{Casagrande:2008hr} for explicit formulae) and that, with appropriately chosen bulk mass parameters $c_i$, one can reproduce the correct values for the masses of the SM quarks. The dominant corrections to the $gg\to h$ and $h\to\ga\ga$ amplitudes arise from fermionic loop contributions and scale with $y_\star^2$ \cite{Casagrande:2010si,Azatov:2010pf,Azatov:2011qy,Goertz:2011hj,Carena:2012fk,Malm:2013jia}. The value of $y_\star$ should be naturally of $\ord(1)$, and requiring that one-loop corrections to the Yukawa couplings remain perturbative one can derive an upper bound $y_\star\lesssim 3$ \cite{Csaki:2008zd} (see also \cite{Malm:2013jia} for a detailed discussion). The green, red, and blue scatter points in the figure correspond to RS model points obtained using three different values of $y_\star$. The latest experimental values for $R_{\gamma\gamma}$ reported by the ATLAS and CMS collaborations are $R_{\gamma \gamma}^{\rm ATLAS}=1.55^{+0.33}_{-0.28}$ (at $m_h= 125.5$\,GeV)  \cite{Aad:2013wqa} and $R_{\gamma\gamma}^{\rm CMS}=0.77\pm 0.27$ (at $m_h=125.7$\,GeV) \cite{CMSresult} which we naively average to obtain $R_{\gamma\gamma}=1.08^{+0.21}_{-0.19}$. The $1\sigma$ error band corresponding to this result is shown by the blue band in the two plots. Model points falling outside these bands are excluded at 68\% confidence level (CL). It is interesting to observe that for relatively large values for $y_\star$ the data already disfavor KK gluon masses in the low TeV range. The tensions between our theoretical predictions and the experimental data are stronger for the brane-Higgs model due to the mild tendency of an enhanced cross section seen in the data, which is in conflict with the suppression of the predicted cross section in this case. We emphasize, however, that using the individual values for $R_{\ga\ga}$ reported by ATLAS and CMS one would obtain different conclusions.

The shape of the various bands of scatter points shown in the plots can be understood as follows. For not too small Yukawa couplings, the largest RS corrections are those arising from fermionic loop contributions. In the brane-localized Higgs (narrow bulk-Higgs) scenario, they suppress (enhance) the gluon-fusion cross section and enhance (suppress) the decay rate into photons. Since the dominant SM contribution to $h\to\ga\ga$ involves $W$-boson loops and acts in the opposite direction as the fermionic contributions, the RS corrections to the Higgs production cross section dominate over those to the decay rate. Hence, we find a suppression (an enhancement) of $R_{\ga\ga}$ in the brane-Higgs (narrow bulk-Higgs) scenario. To see this more explicitly, it is instructive to expand the various expressions in (\ref{eqn:Rgaga}) to first order in $v^2/\mkk^2$ and to approximate $A_q(\tau_t)\approx 1$ and $A_q(\tau_b)\approx 0$. Keeping the dependence on $A_W(\tau_W)\approx 1.19$ explicit, we obtain
\begin{eqnarray}\label{eqn:RgaMin}
   R_{\ga\ga} &\approx& 1 + \frac{v^2}{2\mkk^2} \Bigg[ 
    \left( f_{\rm GF} - \frac{4}{3|C_\gamma^{\rm SM}|} \right) 
    \left( \mp 18 - \frac{10}{3} \right) y_\star^2 \nonumber\\
   &&\hspace{2.05cm}\mbox{}- \left( f_{\rm VBF} 
    + \frac{21\big[ A_W(\tau_W) - 1 \big]}{4|C_\gamma^{\rm SM}|} \right)
    \frac{2m_W^2}{v^2} \left( L - 1 + \frac{1}{2L} \right) \\
   &&\hspace{2.05cm}\mbox{}- \frac{L m_W^2}{v^2} + 0.57\,\frac{10}{3}\,y_\star^2 
    + 0.22\,\frac{2m_W^2}{v^2} \left( L - 1 + \frac{1}{2L} \right)
    - 0.09 \left( \mp 18 - \frac{10}{3} \right) y_\star^2 \Bigg] \,, \nonumber
\end{eqnarray}
where the first two lines contain the corrections to the production and decay rates, with corrections to the $h\to\ga\ga$ rate being accompanied by a factor of $1/|C_\gamma^{\rm SM}|$ with $C_\gamma^{\rm SM}\approx\frac43-\frac{21}{4} A_W(\tau_W)\approx-4.9$. The third line shows the corrections to the Higgs vev and total width, as parameterized by $\kappa_h$ in (\ref{eqn:kh}). The upper sign holds for the brane-localized Higgs scenario, while the lower sign corresponds to the narrow bulk-Higgs case. Above we have used that for a large set of random complex matrices on average \cite{Malm:2013jia}
\begin{equation}
   \left\langle \tr\,\bm{Y}_f\bm{Y}_f^\dagger \right\rangle 
   = N_g^2\,\frac{y_\star^2}{2} \,, \qquad
   \bigg\langle \frac{\left( \bm{Y}_u\bm{Y}_u^\dagger\bm{Y}_u \right)_{33}}%
                     {\left( \bm{Y}_u \right)_{33}} \bigg\rangle 
   = (2N_g-1)\,\frac{y_\star^2}{2} \,,
\end{equation}
where $N_g=3$ is the number of fermion generations. We explicitly see from the first term on the right-hand side of (\ref{eqn:RgaMin}) that the fermionic contributions to the $gg\to h$ production process dominate over those to the $h\to\ga\ga$ decay rate and come with opposite sign. Altogether, we find
\begin{equation}\label{sumup}
   R_{\ga\ga} \approx 1 - \frac{v^2}{2\mkk^2} \big[ (\pm 9.7 - 0.1 )\,y_\star^2 + 4.1 \big] \,.
\end{equation}
For the case where $y_*=3$ this result is shown by the dashed lines in the figure. Note also that due to the contribution of the VBF production process the observable $R_{\gamma\gamma}$ is bounded from below in the brane-Higgs case. This explains the behavior for very small $\mkk$ values seen in the left plot in Figure~\ref{fig:RggMRS}. For $y_*=3$, the $gg\to h$ production cross section vanishes for $M_{g^{(1)}}\approx 3.5$\,TeV, because the new-physics contribution cancels the SM amplitude. However, due to the VBF production process a non-zero value of $R_{\ga\ga}$ remains. 

\begin{figure}
\begin{center}
\psfrag{x}[]{\small $M_{g^{(1)}}~{\rm [TeV]}$}
\psfrag{y}[b]{\small $y_\star$}
\psfrag{z}[]{\small \begin{tabular}{l} minimal RS model \\[-1mm] brane Higgs \end{tabular}}
\psfrag{w}[]{\small \begin{tabular}{l} minimal RS model \\[-1mm] bulk Higgs \end{tabular}}
\includegraphics[width=1.03\textwidth]{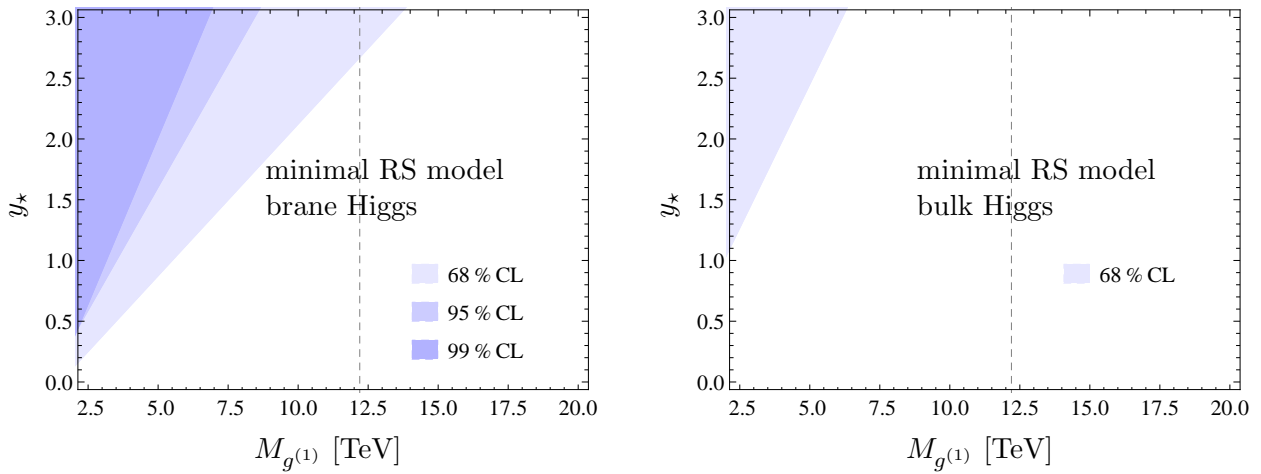}
\parbox{15.5cm}
{\caption{\label{fig:RggExcMRS}
Excluded regions of parameter space in the minimal RS model, for the brane-localized Higgs scenario (left) and the narrow bulk-Higgs model (right). The vertical dashed line denotes the lower bound on $M_{g^{(1)}}$ obtained from a tree-level analysis of electroweak precision observables \cite{Carena:2003fx}.}}
\end{center}
\end{figure}

Even at the present level of precision, the existing measurements of the observable $R_{\ga\ga}$ already provide some interesting constraints on the parameter space of the RS models under consideration. In Figure~\ref{fig:RggExcMRS} we show the regions in the $M_{g^{(1)}}$\,--\,$y_\star$ parameter space that are excluded by the current data at various confidence levels. For instance, for the particular choice $y_\star=3$ one finds $M_{g^{(1)}}>8.5$\,TeV at 95\% CL for the brane-Higgs model and $M_{g^{(1)}}>6.4$\,TeV at 68\% CL for the narrow bulk-Higgs model. Weaker constraints are obtained for smaller values for $y_\star$. These bounds can be compared with the constraints derived from a tree-level analysis of the electroweak $S$ and $T$ parameters \cite{Peskin:1991sw}, which in the minimal RS model leads to the lower bound $M_{g^{(1)}}>12$\,TeV at 95\% CL \cite{Carena:2003fx}. This value is indicated by the vertical dashed line in the figure. At present, this bound is still stronger than the constraints derived from $R_{\gamma\gamma}$. 

\begin{figure}
\begin{center}
\psfrag{x}[]{\small $M_{g^{(1)}}~{\rm [TeV]}$}
\psfrag{y}[b]{\small $R_{\gamma \gamma}$}
\psfrag{z}[]{\small \begin{tabular}{l} custodial RS model \\[-1mm] brane Higgs \end{tabular}}
\psfrag{w}[]{\small \begin{tabular}{l} custodial RS model \\[-1mm] bulk Higgs \end{tabular}}
\psfrag{a}[bl]{\footnotesize $~y_\star=0.5$}
\psfrag{b}[bl]{\footnotesize $~y_\star=1.5$}
\psfrag{c}[bl]{\footnotesize $~y_\star=3$}
\includegraphics[width=1.03\textwidth]{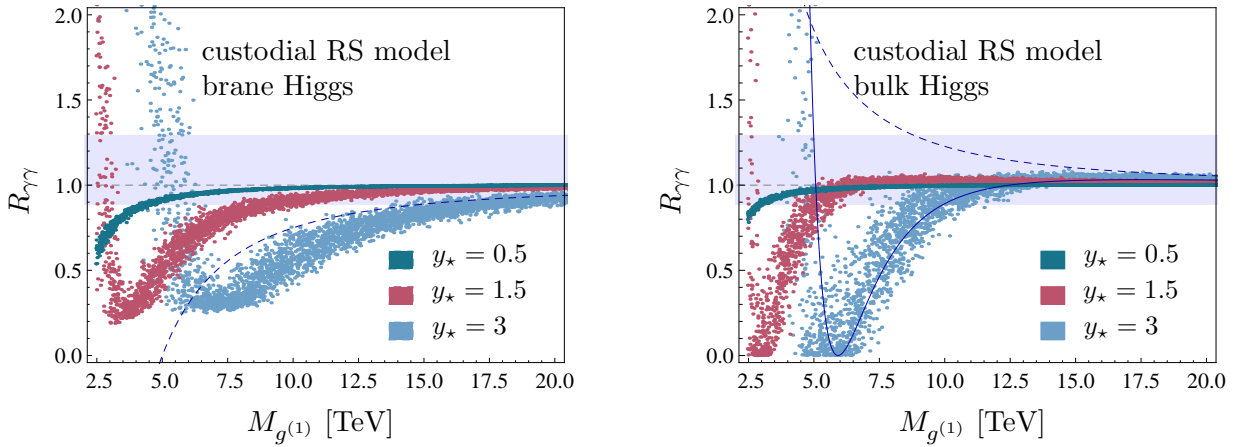}
\parbox{15.5cm}
{\caption{\label{fig:RggCRS}
Predictions for the ratio $R_{\gamma\gamma}$ as a function of the KK gluon mass $M_{g^{(1)}}$ in the custodial RS model with minimal lepton sector (\ref{lepton2}), for the cases of a brane-localized Higgs boson (left) and a narrow bulk-Higgs field (right).}}
\end{center}
\end{figure}

Softening the constraints from electroweak precision observables by means of a symmetry has been the main motivation for extending the RS model by enlarging the bulk gauge group \cite{Agashe:2003zs,Csaki:2003zu,Agashe:2006at}. It has been pointed out in \cite{Goertz:2011hj,Malm:2013jia} that the large number of heavy fermionic degrees of freedom in such an extended model can potentially give rise to large virtual effects on the Higgs-boson production and decay rates. The corresponding effects on the quantity $R_{\ga\ga}$ arising in the RS model with custodial symmetry are studied in Figure~\ref{fig:RggCRS}. In analogy with \eqref{eqn:RgaMin}, we can expand the result in powers of $v^2/\mkk^2$, exploiting the anarchy of the 5D Yukawa matrices and making the same approximations as above. For the model with the minimal lepton sector shown in (\ref{lepton2}), this yields
\begin{equation}\label{eqn:RgaCust}
\begin{aligned}
   R_{\ga\ga} &\approx 1 + \frac{v^2}{2\mkk^2} \Bigg[ 
    \mp \left( 72 f_{\rm GF} - \frac{213}{|C_\gamma^{\rm SM}|} \right) y_\star^2
    - \frac{20}{3} \left( f_{\rm GF} - \frac{4}{3|C_\gamma^{\rm SM}|} \right) y_\star^2 \\    
   &\hspace{2.6cm}\mbox{}- \left( f_{\rm VBF}
    + \frac{21\big[ A_W(\tau_W) - 1 \big]}{4|C_\gamma^{\rm SM}|} \right)
    \frac{2m_W^2}{v^2} \left( 2L - 1 + \frac{1}{2L} \right) - \frac{2L m_W^2}{v^2} \\
   &\hspace{2.6cm}\mbox{}+ 0.57\,\frac{20}{3}\,y_\star^2 
    + 0.22\,\frac{2m_W^2}{v^2} \left( 2L - 1 + \frac{1}{2L} \right)
    - 0.09 \left( \mp 72 - \frac{20}{3} \right) y_\star^2 \Bigg] \,.
\end{aligned}
\end{equation}
If instead the extended lepton sector shown in (\ref{lepton1}) is employed, then the coefficient 213 inside the parenthesis in the first term must be replaced by 240. Note that the individual corrections due to fermion loops are huge, however significant cancellations take place when one adds the corrections to the $gg\to h$ and $h\to\ga\ga$ rates. Altogether, we obtain for the model with the minimal lepton sector
\begin{equation}
   R_{\ga\ga} \approx 1 - \frac{v^2}{2\mkk^2} \big[ (\pm 15.0 - 0.2 )\,y_\star^2 + 8.3 \big] \,.
\end{equation}
In the model with the extended lepton sector the coefficient $\pm 15.0$ in the first term must be replaced by $\pm 9.5$. We observe that in linearized form the corrections are only moderately larger than in the minimal model, see (\ref{sumup}). Once again, for $y_*=3$ this result is shown by the dashed lines in the figure, where we show results for the custodial model with the minimal lepton sector. If instead the model with an extended lepton sector is considered, the distribution of scatter points looks very similar. For the brane-localized Higgs scenario, Figure~\ref{fig:RggCRS} shows a similar behavior as in the minimal model, but the new-physics effects are slightly larger. For $y_*=1.5$ and~3, the $gg\to h$ production cross section vanishes near $M_{g^{(1)}}\approx 3.5$\,TeV and 7\,TeV, respectively, and the VBF process remains as the only production mechanism. This explains the minimum values for $R_{\ga\ga}$ at these points. For even smaller masses the ratio $R_{\ga\ga}$ increases and can even exceed~1. In the narrow bulk-Higgs case, on the other hand, the linearized approximation (\ref{eqn:RgaCust}) breaks down for large values $y_*$, as is evident from the discrepancy between the dashed curve and the blue band of scatter points. A reasonable approximation, shown by the solid line, is obtained by linearizing the expressions for the various $\kappa_i$ parameters but not further expanding expression (\ref{eqn:Rgaga}). It turns out that the negative corrections to the $h\to\ga\ga$ decay rate are so significant in this model that they compensate the large positive corrections to the gluon-fusion rate in the region of large $M_{g^{(1)}}$. For smaller KK masses, these negative corrections become dominant and drive the ratio $R_{\ga\ga}$ toward values significantly less than~1. Eventually, for $M_{g^{(1)}}\approx 3$\,TeV (for $y_*=1.5$) and 5.5\,TeV (for $y_*=3$), the di-photon decay rate even vanishes. It is obvious that in regions of parameter space where such dramatic cancellations occur our predictions are highly model dependent. Given the preliminary pattern of Higgs couplings seen in experiment, which within errors agree with the SM predictions, it appears unlikely (but not impossible) that there could be $\ord(1)$ corrections to the $gg\to h$ and $h\to\ga\ga$ production and decay rates, which cancel each other in the result for the observable $R_{\ga\ga}$. Too large corrections to the gluon-fusion rate are also disfavored by the good agreement of the $pp\to ZZ^{(*)}\to 4l$ rate with its SM value. A detailed discussion of the corresponding constraints on the RS parameter space has been presented in \cite{Malm:2013jia}.

\begin{figure}
\begin{center}
\psfrag{x}[]{\small $M_{g^{(1)}}~{\rm [TeV]}$}
\psfrag{y}[b]{\small $y_\star$}
\psfrag{z}[]{\small \begin{tabular}{l} custodial RS model \\[-1mm] brane Higgs \end{tabular}}
\psfrag{w}[]{\small \begin{tabular}{l} custodial RS model \\[-1mm] bulk Higgs \end{tabular}}
\includegraphics[width=1.03\textwidth]{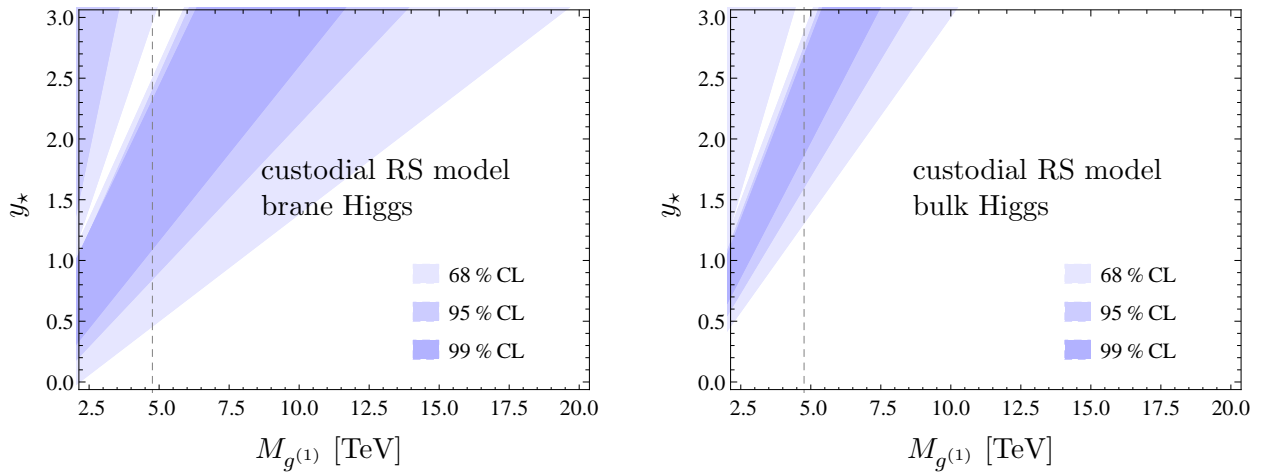}
\parbox{15.5cm}
{\caption{\label{fig:RggExcCRS}
Excluded regions of parameter space in the custodial RS model with minimal lepton sector (\ref{lepton2}), for the brane-localized Higgs scenario (left) and the narrow bulk-Higgs model (right). The vertical dashed line denotes the lower bound on $M_{g^{(1)}}$ obtained from a tree-level analysis of electroweak precision observables \cite{Carena:2003fx}.}}
\end{center}
\end{figure}

Figure~\ref{fig:RggExcCRS} shows the excluded regions of RS parameter space derived from the analysis of the observable $R_{\ga\ga}$ in the custodial RS model. In the scenario with a brane-localized Higgs sector, we obtain the exclusion range $5.9\,{\rm TeV}<M_{g^{(1)}}<13.4$\,TeV and $M_{g^{(1)}}<3.5$\,TeV for $y_\star=3$, while in the narrow bulk-Higgs model we can exclude $5.2\,{\rm TeV}<M_{g^{(1)}}<8.4$\,TeV, both at 95\% CL. Note that there is a small region in the upper left corner (at small $M_{g^{(1)}}$ and large $y_\star$) of the left plot, which is allowed by both $R_{\gamma\gamma}$ and the $T$ parameter constraint $M_{g^{(1)}}>4.7$\,TeV at 95\% CL. However, bounds derived from the analysis of the decay $h\to ZZ^{(*)}\to 4l$ exclude this region \cite{Malm:2013jia,phenopaper}. 

One can also read the exclusion plots in Figures~\ref{fig:RggExcMRS} and \ref{fig:RggExcCRS} in a different way. Under the optimistic assumption that the first KK gluon resonance is in reach for direct production at the LHC, these plots allow one to impose bounds on $y_\star$. For instance, in the minimal RS model with a hypothetical KK gluon mass $M_{g^{(1)}}=5$\,TeV, our results imply an upper bound of $y_\star<1.5$ at 95\% CL in the brane-Higgs model, and $y_\star<2.4$ at 68\% CL in the narrow bulk-Higgs scenario. In the custodial RS model, those bounds are tightened to $y_\star<0.9$ for a brane Higgs and $y_\star<1.7$ for a narrow bulk Higgs, both at 95\% CL. Even though the constraints are rather strong in the case of the custodial RS model, they do not quite compete with those stemming from the decays $h\to ZZ^{(*)}, WW^{(*)}$ \cite{Malm:2013jia,phenopaper}. This is due to the fact that the RS corrections to the decay into two photons partially compensate the huge effect in the gluon-fusion production process. This compensation does not occur in the decays into two weak gauge bosons, whose couplings to the Higgs are only slightly affected by new-physics effects.

\section{Conclusions}
\label{sec:conclusions}

The discovery of a Higgs boson at the LHC in the summer of 2012 \cite{ATLAS:2012gk,CMS:2012gu} has marked the beginning of a new era in elementary particle physics. The couplings of this new particle are found to be non-universal and indeed very close to those predicted for the elementary scalar boson of the SM. An explanation for the hierarchy problem is thus more pressing than ever. Measuring precisely the Higgs couplings to various SM particles provides an important tool to discover and distinguish between new-physics models that can address the hierarchy problem. Especially interesting are loop-induced processes like Higgs production via gluon fusion $gg\rightarrow h$ and the radiative decay $h\rightarrow \gamma\gamma$. 

In this paper, we have focused on the Higgs decay into two photons in different scenarios of RS models, where the Higgs sector is localized on or near the IR brane, while the remaining fields propagate in the bulk of the warped extra dimension. While the contribution from diagrams with virtual fermions (quarks and charged leptons) in the loop has been extensively discussed in recent works, mostly in the context of gluon fusion \cite{Casagrande:2010si,Azatov:2010pf,Azatov:2011qy,Goertz:2011hj,Carena:2012fk,Malm:2013jia}, here we have further concentrated our analysis on the diagrams with bosonic fields propagating in the loops, generalizing the findings of \cite{Cacciapaglia:2009ky,Bhattacharyya:2009nb,Bouchart:2009vq,Casagrande:2010si,Azatov:2010pf}. We have shown that the relevant diagrams, calculated in the general $R_\xi$ gauge, add up to a gauge-invariant result. Working in unitary gauge, we have derived an exact expression for the $h\to\ga\ga$ decay amplitude in terms of an integral over the 5D $W$-boson propagator with the Higgs-boson profile along the extra dimension, given in \eqref{eqn:C1W}. This expression can be used to calculate the bosonic contributions to the amplitude as long as one is able to derive an analytic expression for the 5D $W$-boson propagator. We have shown that the 5D loop diagrams with bosonic fields are insensitive to the precise details of the localization of the scalar sector on or near the IR brane. This finding is in contrast to the case of fermionic loops, for which one finds different results for the cases of a strictly localized Higgs sector and a scenario in which the Higgs is a narrow bulk field \cite{Carena:2012fk,Malm:2013jia}. In our approach we retain the exact dependence of the amplitude on the Higgs-boson mass. Taking a 5D perspective on the calculation, we have not distinguished between the SM modes and the KK particles. However, in our final results we have been able to identify the contributions from the $W$ boson and the infinite tower of its KK excitations. Finally, we have generalized our findings to the RS model with custodial protection of electroweak precision observables, for which we have discussed two different embeddings of the lepton sector in the extended gauge group of the model. 
 
In the phenomenological part of this paper, we have analyzed the new-physics effects on the cross section times branching ratio for the process $pp\to h\to\gamma\gamma$, including effects on the production cross sections via gluon fusion or vector-boson fusion, the $h\to\gamma\gamma$ decay rate, and the total Higgs width. We have focused on the ratio $R_{\ga\ga}$ representing the cross section times branching ratio in the RS model normalized to its SM value, and distinguished between the brane-localized and the narrow bulk-Higgs scenarios of both the minimal and the custodial RS models. We have also derived approximate formulas for $R_{\ga\ga}$, which allow for a reasonable parametrization of our results in terms of the KK mass scale and the maximal value $y_\star$ imposed on the magnitude of the individual entries of the anarchic 5D Yukawa matrices. We have pointed out the fact that the RS corrections to Higgs production via gluon fusion act in the opposite direction as those to the decay into two photons, where generally the new-physics effects on the gluon-fusion rate provide the dominant source of corrections to the SM predictions. For not too small values of $y_\star$ and not too large values of $\mkk$, new-physics effects in RS models can lead to significant deviations of $R_{\ga\ga}$ from~1. While in all brane-Higgs scenarios considered here $R_{\ga\ga}<1$ in most regions of parameter space (in good agreement with the CMS data), RS scenarios with the Higgs implemented as a narrow bulk field allow for the possibility that $R_{\ga\ga}>1$ (as suggested by the ATLAS data). More precise measurements of $R_{\ga\ga}$ and of other Higgs production and decay rates would be required to differentiate between different incarnations of RS models. Importantly, significant deviations of the Higgs couplings from their SM values could arise even if the KK mass scale is so large that KK modes are out of reach for direct production at the LHC. 

Comparing our results with the latest ATLAS and CMS data \cite{Aad:2013wqa,CMSresult}, we have derived exclusion regions in the $M_{g^{(1)}}$\,--\,$y_\star$ parameter space of the various models, shown in Figures~\ref{fig:RggExcMRS} and~\ref{fig:RggExcCRS}. In the minimal model, we can exclude KK gluon masses lighter than $8.5\,\TeV\times (y_\star/3)$ at 95\% CL for the brane-localized scenario, and $6.4\,\TeV\times(y_\star/3)$ at 68\% CL for the narrow bulk-Higgs scenario. In case of the custodially protected RS model and for a brane-localized Higgs, we can exclude KK masses below $13.4\,\TeV\times (y_\star/3)$ at 95\% CL. For a narrow bulk-Higgs field, we find the excluded region $M_{g^{(1)}}<8.4\,\TeV\times (y_\star/3)$ at 95\% CL. 

Together with our previous work \cite{Malm:2013jia}, the present paper allows for a full treatment of the loop-mediated Higgs couplings and can be supplemented by the remaining tree-level couplings to arrive at a comprehensive description of Higgs physics in RS models, where the scalar sector lives on or near the IR brane. With increasing experimental precision on the extracted Higgs couplings, it will be exciting to compare our results for various RS models with the data. Even if no KK excitations will be discovered at the LHC, it is conceivable that future precision measurements of Higgs couplings at the LHC and ILC could provide a first hint on the existence of a warped extra dimension.

\vspace{3mm}
{\em Acknowledgements:\/}
We are grateful to Paul Archer for useful discussions. This research is supported by the Advanced Grant EFT4LHC of the European Research Council (ERC), the Cluster of Excellence {\em Precision Physics, Fundamental Interactions and Structure of Matter\/} (PRISMA -- EXC 1098), grant 05H12UME of the German Federal Ministry for Education and Research (BMBF), the Rhineland-Palatinate Research Center {\em Elementary Forces and Mathematical Foundations}, and the DFG Graduate School GRK~1581 {\em Symmetry Breaking in Fundamental Interactions}. With the exception of C.H., the authors gratefully acknowledge KITP Santa Barbara for hospitality and support during the early stages of this work. KITP is supported by the National Science Foundation under Grant No.\ NSF PHY11-25915. 

\newpage
\begin{appendix}

\section{Feynman rules in the 4D effective theory}
\label{app:FR}
\renewcommand{\theequation}{A.\arabic{equation}}
\setcounter{equation}{0}

\begin{figure}[h]
\begin{center}
{\small
\begin{tikzpicture}[
scale=0.35,>=latex, level/.style={thick}, scalarh/.style={thick,densely dashed,postaction={decorate},decoration={markings,mark=at position .55 with {\arrow[draw=black]{>}}}},scalar2/.style={thick,densely dashed,postaction={decorate},decoration={markings,mark=at position .55 with {\arrow[draw=black]{>}}}}
,scalar1/.style={thick,densely dashed,postaction={decorate},decoration={markings,mark=at position .55 with {\arrow[draw=black]{>}}}},scalar/.style={thick,densely dashed,postaction={decorate}},vector/.style={decorate, decoration={snake}}, pile/.style={thick, ->}, ghost/.style={thick,dotted}]
\begin{scope}[xshift= - 14cm, yshift=0cm]
\node at (-2,3) {$(a)$};
\draw[scalar] (0,0)node[left] {$h$} -- node[above]{}(3.5, 0);
\draw[scalar](3.5,0)--node[above left]{}(5.5,2)node[right]{$\varphi^{\pm(n)}$};
\draw[vector](3.5,0)--(5.5,-2)node[right]{$W_{\nu}^{\mp(m)}$};
\draw[pile] (4.6,2) -- node[above left]{$p_\varphi$}(3.5,1);
\draw[pile] (1,-0.5) -- node[below]{$p_h$}(2.5,-0.5);
\end{scope}
\begin{scope}[xshift=0cm, yshift=0cm]
\node at (-2,3) {$(b)$};
\draw[scalar] (0,0)node[left] {$h$} -- (3.5, 0);
\draw[vector](3.5,0)--(5.5,2)node[right]{$W_{\mu}^{\pm(n)}$};
\draw[vector](3.5,0)--(5.5,-2)node[right]{$W_{\nu}^{\mp(m)}$};
\end{scope}
\begin{scope}[xshift=14cm, yshift=0cm]
\node at (-2,3) {$(c)$};
\begin{scope}[xshift=2cm]
\draw[vector] (2,0) -- (4, -2)node[right]{$A_{\mu}^{(0)}$};
\draw[vector](2,0)--(4,2)node[right]{$W_{\nu}^{\mp(m)}$};
\draw[scalar](0,2)node[left] {$h$}--(2,0);
\draw[scalar](0,-2)node[left] {$\varphi^{\pm(n)}$}--(2,0);
\end{scope}
\end{scope}
\begin{scope}[xshift= - 14cm, yshift=-7cm]
\node at (-2,3) {$(d)$};
\draw[scalar] (0,0)node[left] {$h$} -- (3.5, 0);
\draw[scalar](3.5,0)--(5.5,2)node[right]{$\varphi^{+(n)}$};
\draw[scalar](3.5,0)--(5.5,-2)node[right]{$\varphi^{-(m)}$};
\end{scope}
\begin{scope}[xshift= 0cm, yshift=-7cm]
\node at (-2,3) {$(e)$};
\begin{scope}[xshift=-1cm]
\draw[vector] (5,0) -- (7, -2)node[right]{$A_{\mu}^{(0)}$};
\draw[vector](5,0)--(7,2)node[right]{$A_{\nu}^{(0)}$};
\draw[vector](3,2)node[left] {$W_{\lambda}^{+(n)}$}--(5,0);
\draw[vector](3,-2)node[left] {$W_{\rho}^{-(m)}$}--(5,0);
\end{scope}
\end{scope}
\begin{scope}[xshift= 14cm, yshift=-7cm]
\node at (-2,3) {$(f)$};
\begin{scope}[xshift=2cm]
\draw[vector] (2,0) -- (5.5, 0)node[right]{$A_{\mu}^{(0)}$};
\draw[scalar](0,2)node[left] {$\varphi_W^{+(n)}$}--node[above right]{}(2,0);
\draw[scalar](0,-2)node[left] {$\varphi_W^{-(m)}$}--node[below right]{}(2,0);
\draw[pile] (1,2) -- node[above right]{$p_+$}(2,1);
\draw[pile] (1,-2) -- node[below right]{$p_-$}(2,-1);
\end{scope}
\end{scope}
\begin{scope}[xshift= - 14cm, yshift=-14cm]
\node at (-2,3) {$(g)$};
\begin{scope}[xshift=6cm]
\draw[vector] (-2,0)-- (-5.5, 0)node[left] {$A_{\mu}^{(0)}$} ;
\draw[scalar](0,2)node [right] {$\varphi_W^{\pm(m)}$}--(-2,0);
\draw[vector](0,-2)node [right] {$W_{\nu}^{\mp(n)}$}--(-2,-0);
\end{scope}
\end{scope}
\begin{scope}[xshift= 0cm, yshift=- 14cm]
\node at (-2,3) {$(h)$};
\begin{scope}[xshift=2cm]
\draw[vector] (2,0) -- (4, -2)node[right]{$A_{\mu}^{(0)}$};
\draw[vector](2,0)--(4,2)node[right]{$A_{\nu}^{(0)}$};
\draw[scalar](0,2)node[left] {$\varphi_W^{+(n)}$}--(2,0);
\draw[scalar](0,-2)node[left] {$\varphi_W^{-(m)}$}--(2,0);
\end{scope}
\end{scope}
\begin{scope}[xshift=  14cm, yshift=-14cm]
\node at (-2,3) {$(i)$};
\begin{scope}[xshift=2cm]
\draw[vector] (2,0) -- (5.5, 0)node[right]{$A_{\mu}^{(0)}$};
\draw[ghost] (0,2)node[left]{$c_{\pm}^{(n)}$}--node[above right]{}(2,0);
\draw[thick,dotted](0,-2)node[left] {$c_{\mp}^{(m)}$}--(2,0);
\draw[pile] (2,1) -- node[above right]{$p_\mu$}(1,2);
\end{scope}
\end{scope}
\begin{scope}[xshift= -14cm, yshift=-21cm]
\node at (-2,3) {$(j)$};
\draw[scalar] (0,0)node[left]{$h$} -- (3.5, 0);
\draw[ghost](3.5,0)--(5.5,2)node[right]{$c_{\mp}^{(n)}$};
\draw[ghost](3.5,0)--(5.5,-2)node[right] {$c_{\pm}^{(m)}$};
\end{scope}
\begin{scope}[xshift=0cm, yshift=-21cm]
\node at (-2,3) {$(k)$};
\begin{scope}[xshift=2cm]
\draw[pile] (4.5,0.5)node[above left]{$p_{\gamma}$} -- (3,0.5);
\draw[pile] (1,2) -- node[above right]{$p_+$}(2,1);
\draw[pile] (1,-2) -- node[below right]{$p_-$}(2,-1);
\draw[vector] (2,0)-- (5.5, 0)node[right] {$A_{\mu}^{(0)}$} ;
\draw[vector](0,2)node[left]{$W_{\rho}^{+(n)}$}--(2,0);
\draw[vector](0,-2)node[left]{$W_{\sigma}^{-(m)}$}--(2,-0);
\end{scope}
\end{scope}
\end{tikzpicture}
}
\end{center}
\end{figure}

\noindent
Here we list the Feynman rules needed for the calculation of the one-loop gauge-boson, scalar, and ghost contributions to the $h\to\ga\ga$ decay amplitude in the KK-decomposed version of the minimal RS model, using the notations of \cite{Casagrande:2008hr}. We work in a general $R_\xi$ gauge and use mass eigenstates of all SM particles and their KK excitations. The Feynman rules for the vertices shown above are:
\begin{equation}
\begin{aligned}
   \mbox{$(a)$:}\quad 
   & \frac{\tilde m_W^2}{v m_n^W}\,2\pi\,\chi_m^W(1)\,\chi_n^W(1)\,(p_\varphi-p_h)_\nu \,, 
   &\qquad
   \mbox{$(b)$:}\quad
   & \frac{2i\tilde m_W^2}{v}\,2\pi\,\chi_m^W(1)\,\chi_n^W(1)\,\eta_{\mu\nu} \,, \\
   \mbox{$(c)$:}\quad
   & \pm e\,\frac{\tilde m_W^2}{v m_n^W}\,2\pi\,\chi_m^W(1)\,\chi_n^W(1)\,\eta_{\mu\nu} \,, 
   &\qquad
   \mbox{$(d)$:}\quad
   & - \frac{im_h^2}{v}\,\frac{\tilde m_W^2}{m_m^W m_n^W}\,2\pi\,\chi_m^W(1)\,\chi_n^W(1) \,, \\
   \mbox{$(e)$:}\quad
   & -ie^2\,\big( 2\eta_{\lambda\rho}\,\eta_{\mu\nu} - \eta_{\lambda\mu}\,\eta_{\rho\nu}
    - \eta_{\lambda\nu}\,\eta_{\rho\mu} \big)\,\delta_{mn} \,, 
   &\qquad
   \mbox{$(f)$:}\quad
   & ie\,(p_+ - p_-)_\mu\,\delta_{mn} \,, \\
   \mbox{$(g)$:}\quad
   & \pm e\,m_m^W\,\eta_{\mu\nu}\,\delta_{mn} \,, 
   &\quad
   \mbox{$(h)$:}\quad
   & 2ie^2\,\eta_{\mu\nu}\,\delta_{mn} \,, \\
   \mbox{$(i)$:}\quad
   & \pm ie\,p_\mu\,\delta_{mn} \,, 
   &\quad
   \mbox{$(j)$:}\quad
   & - \xi\,\frac{i\tilde m_W^2}{v}\,2\pi\,\chi_m^W(1)\,\chi_n^W(1) \,, \\
   \mbox{$(k)$:}\quad
   & ie\,\delta_{mn}\,V_{\rho\mu\sigma}(p_+,p_{\gamma},p_-) \,,
   &&
\end{aligned}
\end{equation}
where $v$ is the Higgs vev in the RS model, the parameter $\tilde m_W$ has been defined in (\ref{varphidecomp}), and the tensor structure $V_{\rho\mu\sigma}$ of the 3-boson vertex has been given in the text after (\ref{eqn:5Damp}).

\section{$\bm{W}$ mass and profile derived from the 5D propagator}
\label{sec:massren}
\renewcommand{\theequation}{B.\arabic{equation}}
\setcounter{equation}{0}

For the analysis of the zero-mode contributions in Section~\ref{sec:AnalZeroMRS} we need to expand the $W$-boson propagator at $t=t'=1$ in powers of $v^2/\mkk^2$. In this context, we also need the relation between the physical $W$-boson mass and the model parameter $\tilde m_W$ in (\ref{varphidecomp}) beyond the leading order. A corresponding formula was derived in \cite{Casagrande:2008hr} by solving the eigenvalue equation for the $W$-boson profiles and extracting the lowest eigenvalue. We now present an alternative approach based on our exact expression for the 5D gauge-boson propagator, which allows us to derive explicit expressions for the $W$-boson mass and profile to any order in $v^2/\mkk^2$.  

Starting from the exact expression \eqref{eqn:BbraneMin}, we perform an expansion in powers of $v^2/\mkk^2$ while keeping $p^2$ and $\tilde m_W^2$ fixed and of order $v^2$. This yields
\begin{equation}\label{eqn:PropExp}
   B_W(t,t';-p^2) = \frac{1}{2\pi}\,
    \frac{-1}{(p^2-\tilde m_W^2) \left[ 1+\Pi(t,t';p^2) \right] + \Sigma(p^2) + i0} \,,
\end{equation}
where 
\begin{equation}
\begin{aligned}
   \Sigma(p^2) &= \frac{\tilde m_W^4}{2\mkk^2} \left( L - \frac{p^2}{\tilde m_W^2}
    + \frac{1}{2L}\,\frac{p^4}{\tilde m_W^4} \right) , \\
   \Pi(t,t';p^2) &= \frac{\tilde m_W^2}{2\mkk^2} \left\{ L t_>^2
    + \frac{p^2}{\tilde m_W^2} \left[ L t_<^2 - t^2 \left( \frac12 - \ln t \right) 
    - t'^2 \left( \frac12 - \ln t' \right) \right]  \right\} ,
\end{aligned} 
\end{equation}
which are valid up to terms of order $v^4/\mkk^4$. The zero of the denominator of the propagator in (\ref{eqn:PropExp}) defines the physical mass $m_W$ of the ground state, and the residue of the pole determines the corresponding product of profile functions $\chi_0^W(t)\,\chi_0^W(t')$. Indeed, for $p^2\approx m_W^2$ we find
\begin{equation}
   B_W(t,t';-p^2) = \frac{1}{2\pi}\,\frac{-Z_2(t,t')}{p^2-m_W^2+i0}
    + \mbox{non-singular terms,}
\end{equation}
where 
\begin{equation}
\begin{aligned}
   m_W^2 &= \tilde m_W^2 - \Sigma(m_W^2)
    = \tilde m_W^2 \left[ 1 - \frac{\tilde m_W^2}{2\mkk^2} 
    \left(L -1 + \frac{1}{2L} \right) + \dots \right] , \\
   Z_2(t,t') &= 1 - \Pi(t,t';m_W^2) - \frac{\pa\Sigma(p^2)}{\pa p^2} \Big|_{p^2 = m_W^2} \\
   &= 1 + \frac{\tilde m_W^2}{4\mkk^2} \left[ 2 - \frac{2}{L}
    + t^2 \left( 1 - 2L - 2\ln t \right) + t'^2 \left( 1 - 2L - 2\ln t' \right) \right] 
    + \dots \,.
\end{aligned}
\end{equation}
These results are valid up to $\ord(v^4/\mkk^4)$ corrections. The relation in the first line is equivalent to (\ref{Wmassrela}). Rewriting the second relation in the form $Z_2(t,t')=2\pi\,\chi_0^W(t)\,\chi_0^W(t')$, we can extract
\begin{equation}
   \chi_0^W(t) = \frac{1}{\sqrt{2\pi}} \left\{ 1 + \frac{\tilde m_W^2}{4\mkk^2}
    \left[ 1 - \frac{1}{L} + t^2 \left( 1 - 2L - 2\ln t \right) \right] + \dots \right\} ,
\end{equation}
which gives the first non-trivial correction to the profile of the $W$ boson \cite{Casagrande:2008hr,Csaki:2002gy}. 

\end{appendix}

\end{document}